\pdfoutput=1
\documentclass[11pt,a4paper]{article}
\usepackage{jheppub}

\usepackage{graphicx}% Include figure files
\usepackage{dcolumn}% Align table columns on decimal point
\usepackage{bm}% bold math
\usepackage{epsfig}
\usepackage{enumerate}

\usepackage[usenames,dvipsnames]{xcolor}

\usepackage{amsmath}
\usepackage{amssymb}

\notoc

\begin{document}

\preprint{ANL-HEP-PR-13-17}

\title{A Supersymmetric Theory of Vector-like Leptons}

\author[a]{Aniket Joglekar,}
\emailAdd{aniket@uchicago.edu}
\author[b,c]{Pedro Schwaller}
\emailAdd{pschwaller@hep.anl.gov}
\author[a,b,d]{ and Carlos E. M. Wagner}
\emailAdd{cwagner@hep.anl.gov}

\affiliation[a]{Enrico Fermi Institute, University of Chicago, Chicago, IL 60637, U.S.A.}
\affiliation[b]{HEP Division, Argonne National Laboratory, 9700 Cass Ave., Argonne, IL 60439, U.S.A.}
\affiliation[c]{Physics Department, University of Illinois at Chicago, Chicago, IL 60607, U.S.A.}
\affiliation[d]{Kavli Institute for Cosmological Physics, University of Chicago, Chicago, IL 60637, U.S.A.}

\abstract{
We study a supersymmetric extension of the vector-like lepton scenario, such that the vacuum instability induced by large lepton Yukawa couplings is lifted by the presence of superpartners at or below the TeV scale. In order to preserve the unification of gauge couplings, we introduce a full $16+\overline{16}$ of SO(10), and determine the maximal possible values for the Yukawa couplings consistent with perturbativity at the GUT scale. We find that the Higgs to diphoton decay rate can be enhanced by up to 50\% while maintaining vacuum stability and keeping the new particle masses above 100~GeV, while larger enhancements are possible if the masses of the new particles are lowered further. 
}

\maketitle

%%%%%%%%%%%%%%%%%%
\section{Introduction}\label{sec:into}

In Ref.~\cite{ASW} we have presented a model where the Higgs decay rate to diphotons is increased through loops of mixed vector-like leptons. A vector-like doublet and a vector-like singlet allow for both Yukawa and Dirac masses. The resulting mixing leads to a sign flip of the coupling of the lightest lepton to the Higgs, such that constructive interference with the standard model (SM) amplitude for $h\to\gamma\gamma$ is possible, resulting in a diphoton decay rate that is enhanced relative to the SM. Similar models were presented in~\cite{ArkaniHamed:2012kq,An:2012vp}, and further studies have appeared in~\cite{Kearney:2012zi,Davoudiasl:2012ig,Bae:2012ir,Voloshin:2012tv,McKeen:2012av,Lee:2012wz,Arina:2012aj,Batell:2012zw,Davoudiasl:2012tu,Fan:2013qn,Carmona:2013cq,Cheung:2013bn,Feng:2013mea,Englert:2013tya}
\footnote{Effects of vector-like quark multiplets are discussed e.g. in~\cite{Dawson:2012di,Bonne:2012im,Dawson:2012mk,Moreau:2012da,Bertuzzo:2012bt}.}. 

One of the shortcomings of the model~\cite{ASW} is that the additional ${\cal O}(1)$ Yukawa couplings accelerate the downward running of the Higgs quartic coupling, such that vacuum stability is lost at scales around $10$~TeV. This tension between an enhanced di-photon rate and vacuum stability was first noted in~\cite{ASW,ArkaniHamed:2012kq} and further explored in~\cite{Reece:2012gi,Kitahara:2012pb,Chun:2012jw,Carena:2012mw,Huo:2012tw,Kitahara:2013lfa}. 
The goal of the present paper is to provide a UV completion of the model~\cite{ASW}, which guarantees vacuum stability up to very high scales and unification of gauge couplings at roughly $10^{16}$~GeV.  

In supersymmetric models, the scalar potential is fixed by gauge invariance and supersymmetry relations, so the renormalization group evolution (RGE) of the quartic coupling does not lead to a stability problem, provided that it is well behaved at scales below the SUSY breaking scale. It is therefore natural to try to embed the model~\cite{ASW} into a supersymmetric extension of the SM. 

It is well known that within the MSSM the gauge couplings unify at approximately $10^{16}$~GeV with very high accuracy. In order to not destroy the sensible relation between the values of the couplings at the low scale and the beta function coefficients, additional matter fields have to be embedded in complete multiplets of SU(5). A vector-like doublet with leptonic quantum numbers is naturally contained in a $5+\overline{5}$ representation of SU(5), while a corresponding singlet field can emerge from a $10+\overline{10}$. In the language of SO(10) grand unification, these representations can be combined into a $16+\overline{16}$ representation where the additional degrees of freedom have neutrino-like quantum numbers. 

Additional matter with couplings to the Higgs sector will also give a positive contribution to the mass of the lightest Higgs boson in the MSSM~\cite{Babu:2008ge,Martin:2009bg,Graham:2009gy,Endo:2012cc,ArkaniHamed:2012gw}, thus might help reducing the fine tuning problem that is present in the Higgs sector of the MSSM. 

This paper is organized as follows: In Sec.~\ref{sec:model}, we introduce the model and the particle content at the weak scale. The RGE of the gauge and Yukawa couplings are discussed in Sec.~\ref{sec:rge}. The effects of the new particles on the Higgs mass and di-photon decay rate are calculated in section~\ref{sec:higgs}, while in Sec.~\ref{sec:vacuum} the conditions for vacuum stability are derived. In Sec.~\ref{sec:results} we present numerical results before concluding in Sec.~\ref{sec:conclusions}. 

%%%%%%%%%%%%%%%%%%
\section{The Model}\label{sec:model}

Our model is the minimal supersymmetric standard model (MSSM) extended by a $16+\overline{16}$ of SO(10), such that the unification of gauge couplings at the GUT scale is guaranteed. Models with such a particle spectrum were previously studied e.g. in~\cite{Babu:2008ge}\cite{Martin:2009bg}. Below $M_{\rm GUT}$ the SO(10) multiplets are split. We assume that all the additional colored states obtain masses above the TeV scale, and therefore are in agreement with current LHC limits. 

Below the TeV scale, we assume that the only fields beyond the MSSM particle content are the un-colored components of the original $16+\overline{16}$ supermultiplets. In terms of chiral superfields, these are $SU(2)$ doublets $L'_L$ and $\overline{L''_R}$, the singlet fields $\overline{E_R'}$ and $E''_L$ as well as singlet neutrino superfields $\overline{N'_R}$ and $N''_L$ with the quantum numbers indicated in Tab.~\ref{tab:fields}. Our notation is such that the bar extends over the implicit chiral projector, i.e. $\overline{E'_R}$ is a left-handed superfield. 
\begin{table}
\center
\begin{tabular}{| l |c|c|c|c|c|c|}
\hline
Name & $L'_L = \begin{pmatrix} N'_L\\E'_L \end{pmatrix}$ & 
$\overline{E'_R}$ & $\overline{N'_R}$ & 
$\overline{L''_R} = \begin{pmatrix} \overline{E''_R}\\\overline{N''_R} \end{pmatrix}$ & $E''_L$ &  $N''_L$  \\ \hline
Quantum Numbers & (2, -1/2) & (1, 1) & (1, 0) & (2, 1/2) & (1, -1) & (1, 0) \\ \hline
\end{tabular}
\caption{Additional fields below the TeV scale. 
All fields are left-handed superfields, and the quantum numbers specify the transformations of the fields under the SU(2)$\times$U(1) gauge group of the SM. Primes indicate fields coming from the $16$ multiplet, while the double primed fields originate from the $\overline{16}$ multiplet. \label{tab:fields}}
\end{table}
The superpotential is
\begin{align}
	W =& W_{\rm MSSM} - M_L L'_L \overline{L''_R} + M_E E''_L \overline{E'_R} - y'_c L'_L H_d \overline{E'_R} + y''_c \overline{L''_R} H_u E''_L \label{eqn:superpotential}
	\\\notag
	-& y_{n}' L'_L H_u \overline{N'_R}+y_{n}'' \overline{L''_R} H_d N''_L - M_{ij} N_i N_j \,,
\end{align}
where we have neglected the colored fields beyond the MSSM, and $(N_1,N_2) = (\overline{N'_R}, N''_L)$ since the neutrinos can have Majorana mass terms in addition to the Dirac mass terms for $L$ and $E$ fields. As in~\cite{ASW}, we impose a parity symmetry under which the vector-like multiplets are odd, such that mixing with MSSM leptons and sleptons is forbidden. 

Contraction of SU(2) indices is implicit in~(\ref{eqn:superpotential}). In analogy with the leptonic superfields, the Higgs superfields are defined as  $H_u = (H_u^+, H_u^0)^T$ and $H_d = (H_d^0, H_d^-)^T$. Indices are contracted using the anti-symmetric tensor $\epsilon$ with $\epsilon_{12} = 1$, for example $L'_L H_d = (N_L' H_d^- - E_L' H_d^0)$. The sign conventions in~(\ref{eqn:superpotential}) are chosen such that all entries in the charged fermion mass matrix come with a positive sign and all entries in the neutral fermion mass matrix with a negative sign, for convenience. 

Explicitly, the charged fermion mass matrix is given by
\begin{align}\label{eqn:lepmass}
	\frac{1}{2} \left( e_L' \,e_L'' \, \overline{e'_R}\,  \overline{e_R''} \right)  {\cal M}_{E}  
	\begin{pmatrix}
	 e_L'  \\ e_L''\\ \overline{e'_R} \\ \overline{e_R''}
	\end{pmatrix} + {\rm h.c.} \quad
	& \text{ with }
	{\cal M}_E = 
	\begin{pmatrix}
		0&0 & y_c' v_d & M_L \\
		0 & 0  & M_E &  y_c'' v_u \\
		y_c' v_d & M_E & 0 & 0 \\
		M_L &  y_c'' v_u & 0 & 0 
	\end{pmatrix},
\end{align}
where we used lower case letters for the fermionic components of the superfields, and we have introduced the scalar vacuum expectation values $v_u = \langle H_u^0 \rangle$ and $v_d = \langle H_d^0 \rangle$. In the supersymmetric limit, the corresponding slepton mass matrix is simply given by ${\cal M}_{\tilde{E}}^2 = {\cal M}_E^\dagger {\cal M}_E$. The $\mu H_u H_d$ term communicates SUSY breaking to the lepton sector, such that, in a non-trivial Higgs background, the charged slepton mass matrix assumes the form
\begin{align}\label{eqn:slepmass}
	&{\cal M}^2_{\tilde{E}}  = 
	\\ \notag 
	& \begin{pmatrix}
		|y'_c|^2 v_d^2 + |M_L|^2 + m_{e'_L}^2 & y_c'^{*}v_d M_E + M_L^* y_c'' v_u & -y'_c v_u \mu^*  & b_L \\
		M_E^* y_c' v_d + y_c''^* v_u M_L  & |y''_c|^2 v_u^2 + |M_E|^2 + m_{e''_L}^2 & b_E & -y''_c v_d \mu^* \\
		 -y'^*_c v_u \mu& b_E  & |M_E|^2 + |y_c'|^2 v_d^2 + m_{e_R'}^2 & y_c'^* v_d M_L + M_E^* y_c'' v_u \\
		 b_L& -y''^*_c v_d \mu & M_L^* y_c' v_d + y_c''^* v_u M_E & |M_L|^2 + |y_c''|^2 v_u^2 + m_{e_R''}^2 
	\end{pmatrix},
\end{align}
in the basis $\left( \tilde{e}'_L,\, \tilde{e}''_L,\, \tilde{e}'_R,\, \tilde{e}''_R \right)$. The D-term contributions are not explicitly written down since their effects are small, but they are included in our analysis. Note that, in addition to the supersymmetric mass terms, we have allowed for the following bi-linear soft breaking terms in the potential:
\begin{align}
	V_{\rm soft,\ell} & = m_{e'_L}^2 |\tilde{\ell}'_L|^2 + m_{e''_R}^2 |\tilde{\ell}''_R|^2 + m_{e''_L}^2 |\tilde{e}''_L|^2 + m_{e'_R}^2 |\tilde{e}'_R|^2
	+ {\textstyle \frac{1}{2}}(b_L \tilde\ell_L'^\dagger  \tilde\ell_R''  + b_E  \tilde{e}_R'^* \tilde{e}''_L + {\rm h.c.})\,.
\end{align}
The structure of the mass matrices and their effects on the $h\to \gamma\gamma$ rate will be further explored in Sec.~\ref{sec:higgs}. 
It is worth noting that half of the scalar degrees of freedom can be decoupled by increasing $m_{e_L''}^2$ and $m_{e_R''}^2$ to high values. This limit is similar to the light stau scenario~\cite{Carena:2011aa,Carena:2012gp}, with a $2\times 2$ charged slepton mass matrix 
\begin{align}
	{\cal M}^2_{\tilde{E},2\times 2} & = \begin{pmatrix}
		|y'_c|^2 v_d^2 + |M_L|^2 + m_{e'_L}^2 &  -y'_c v_u \mu^* \\
		-y'^*_c v_u \mu&  |M_E|^2 + |y_c'|^2 v_d^2 + m_{e_R'}^2
	\end{pmatrix} .
\end{align}
A similar contribution is obtained when instead the double primed fields $\tilde{e}''_L$ and $\tilde{e}''_R$ are lifted, however in that case the mixing is proportional to $v_d$ and therefore suppressed in the large $\tan\beta$ regime. 

\

Before moving to the next section, let us briefly review existing experimental limits on uncolored charged scalars and fermions. The LEP experiments have searched for such particles, and their results are collected in the particle data booklet (PDG)~\cite{PDG}. 

Limits on additional charged leptons are given explicitly in the PDG. For a charged lepton that decays to $W^\pm \nu$, a limit of 100.8~GeV is quoted, while a limit of 101.9~GeV is given for a charged lepton that is not degenerate with it's corresponding neutrino. However, as discussed in detail in~\cite{ASW}, if a neutrino-like state is close in mass, these limits become invalidated. 

Limits on the scalar partners of standard model leptons strongly depend on the flavor composition of the sleptons, and range from 107~GeV for left-handed selectrons down to 81.9~GeV for staus. As in the fermionic case, most of these bounds are weakened if other neutral states are close by in mass, and none of them can be directly applied to scalar partners of new leptons, as in our model. As absolute lower bound on the mass of the lightest leptons and slepton we therefore impose $m_{\tilde{E}_1} > m_h/2 \approx 62.5$~GeV. To facilitate the discussion in the remainder of the paper, we define two LEP limits:
\begin{itemize}
	\item Conservative LEP limit: $m_{E_1} > 100$~GeV, $m_{\tilde{E}_1} > 90$~GeV,
	\item Optimistic LEP limit: $m_{E_1} > 62.5$~GeV, $m_{\tilde{E}_1} > 62.5$~GeV,
\end{itemize}
with the understanding that a spectrum that satisfies the conservative LEP limit will certainly be in agreement with current experimental bounds, while a spectrum that satisfies the optimistic LEP limit might require additional invisible neutral states to be close in mass in order to not be in conflict with existing searches. It should be noted that such states are present in our model in the form of new neutrinos and sneutrinos. 

The LHC experiments have not yet performed a dedicated search for signatures of vectorlike leptons. Since we assume that both the lightest new lepton and the lightest new slepton are neutral and stable, the leading visible signatures will come from pair production of heavier states that decay to the lightest state emitting a $W$ or $Z$ boson. In particular, for the leptons we can have $p p \to E_1^\pm N_2 \to W^\pm Z N_1 N_1$, and similarly for the sleptons. The resulting trilepton plus missing energy signature is similar to that of MSSM chargino and neutralino searches, for which results are available from both ATLAS~\cite{ATLAS:2013rla} and CMS~\cite{CMS:aro}. These searches exclude chargino masses up to 300~GeV, however only in regions of parameter space where the mass difference between the lightest and the heavier states is larger than the Z-boson mass. For mass differences up to 50~GeV the limit drops to about 170~GeV, and no limit is available if the mass difference is less than 30~GeV. 

Due to the different SU(2) quantum numbers, the production cross section for vectorlike leptons is roughly a factor of two smaller than that for charginos and neutralinos, while the cross section for slepton pair production is suppressed even further. It follows that we can always evade the current LHC limits by requiring that the new leptons and sleptons are either close in mass to the lightest new state, or heavier than about 300~GeV. Spectra with exactly these features are suggested by our results from~\cite{ASW}, which is why we only impose the LEP bounds on the lepton and slepton masses. Note however that the 14~TeV LHC with sufficient luminosity will be able to improve upon the LEP bounds in most regions of parameter space~\cite{Carena:2012gp,ArkaniHamed:2012kq}.

Contributions to the electroweak $S$ and $T$ parameters from vectorlike leptons were studied in~\cite{ASW}, and were found to be in agreement with the existing limits even for very light masses for the new fermions. The main reason for this is that while the Yukawa couplings induce some custodial symmetry breaking, the effect is not too large since there is no color factor and a suppression of order $y_c^{'(')}v/M_{L,E}$ from the vectorlike mass terms. 
The corresponding slepton contributions were calculated e.g. in~\cite{Martin:2009bg}, and are of the same order as the lepton contributions, such that the overall effect of leptons and sleptons should remain in agreement with the data. Finally we assume that the additional quarks and squarks from the SO(10) multiplets have TeV scale vectorlike masses, such that their contributions decouple. Therefore our model will in general be in agreement with electroweak precision constraints, even with lepton and slepton masses close to the LEP bound.

\section{Running of Couplings}\label{sec:rge}
The RGE of the gauge and Yukawa couplings is strongly affected by the presence of additional matter charged under the strong and weak interactions. To simplify the analysis of this section, we assume a common threshold at the TeV scale for the new vector-like states and for all SUSY partners. While ultimately we will be interested in scenarios where some of the vector-like matter is lighter, the effects on the RGE of the gauge couplings are negligible. 
\begin{figure}
\center
\includegraphics{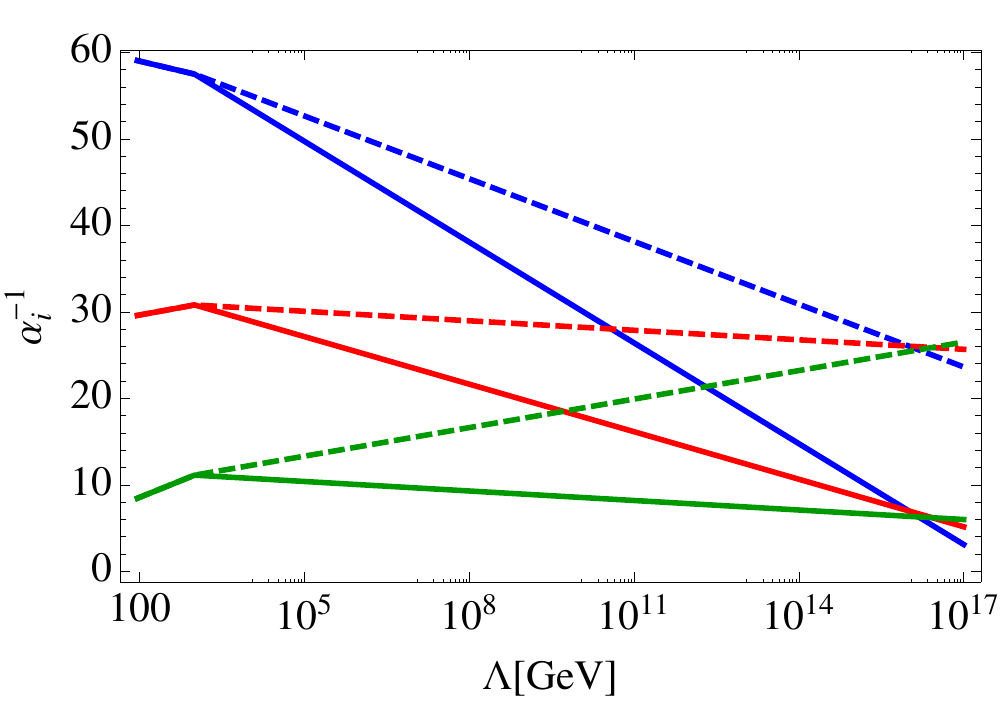}
\caption{Evolution of gauge couplings in the MSSM (dashed lines) and in the MSSM extended by a $16+\overline{16}$ of SO(10) (solid lines). From top to bottom the curves correspond to $\alpha_1$, $\alpha_2$ and $\alpha_3$. }
\label{fig:rgeG}
\end{figure}

The one-loop evolution of gauge couplings is governed by
\begin{align}
	\alpha_i^{-1}(\Lambda) = \alpha_i^{-1} (M_Z) - \frac{b_i}{2\pi} \log \frac{\Lambda}{M_Z}\,.
\end{align}
When $\Lambda > 1$~TeV, the one loop beta function coefficients are given by 
\begin{align}
	\beta_1 & = -\frac{3}{5}-2 N_f =  -\frac{53}{5}\,,\notag\\
	\beta_2 & =  5 -2 N_f = -5 \,, \notag \\
	\beta_3 & =  9 - 2 N_f =  -1\,, \notag
\end{align}
with $N_f=5$, while the corresponding values in the MSSM are obtained by taking $N_f=3$. Note in particular that this leads to a sign change in the beta function for $\alpha_3$, which shows that the full particle content of the MSSM $+16+\overline{16}$ is enough to render the strong interactions asymptotically non-free. 

Fig.~\ref{fig:rgeG} shows the evolution up to high scales. Unification of gauge couplings occurs roughly around $M_U\sim 1.5\times 10^{16}$~GeV, with coupling strengths of about $\alpha_i \sim 0.15$ corresponding to $g_i \sim 1.4$. At the one loop level this is compatible with perturbative unification. For a discussion of higher order effects see e.g.~\cite{Martin:2009bg}. 

Yukawa couplings have a well known fixed point behavior in the infrared, which allows us to determine an upper limit on the magnitude of the couplings at the electroweak scale. At the one-loop level, the RGE equations for the Yukawa couplings take the form
\begin{align}
	\frac{d}{d\mu} y_i(\mu) & = \frac{1}{\mu} b_i(\mu)\,,
\end{align}
with beta functions given by~\cite{Auberson:1999kv}
\begin{align}
	b_t(\mu) & = \frac{1}{16\pi^2} y_t \left( 6 y_t^2 +y_c''^2 - 4 \pi \left( {\textstyle \frac{13}{15}} \alpha_1 + 3 \alpha_2+ {\textstyle \frac{16}{3} } \alpha_3\right)      \right) , \\
	b_{y_c'}(\mu) & =  \frac{1}{16\pi^2} y_c' \left( 4 y_c'^2  - 4 \pi \left( {\textstyle \frac{9}{5}} \alpha_1 + 3 \alpha_2 \right)      \right), \\
	b_{y_c''}(\mu) & =  \frac{1}{16\pi^2} y_c'' \left( 4 y_c''^2 +3y_t^2 -  4 \pi \left( {\textstyle \frac{9}{5}} \alpha_1+ 3 \alpha_2 \right) \right) ,
\end{align}
where we have suppressed the scale dependence on the right-hand sides. For sufficiently large initial values at the weak scale, $y_c'$ and $y_c''$ will diverge at high scales. The top Yukawa coupling is $y_t(M_Z) = M_t/(v \sin\beta)$, such that the bound on $y_c''$ will be more stringent for small $\tan\beta$, while for larger $\tan\beta$ the bottom and tau Yukawas will lead to a slightly stronger bound on $y_c'$. 
\begin{figure}
\center
\includegraphics[width=.48\textwidth]{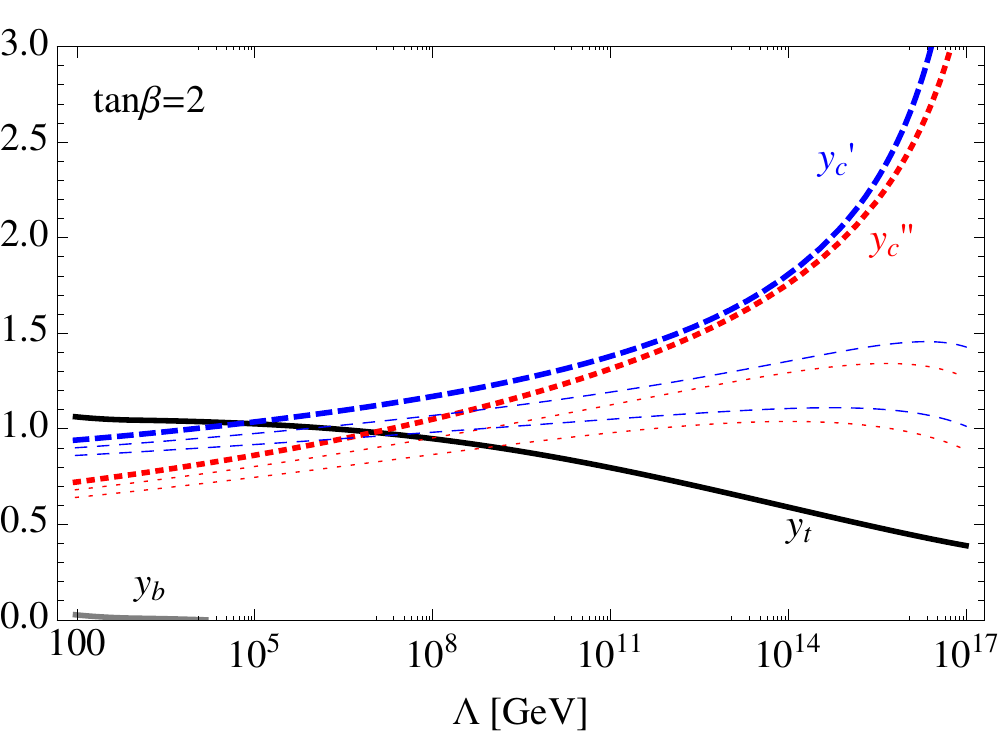}\hspace*{.5cm}
\includegraphics[width=.48\textwidth]{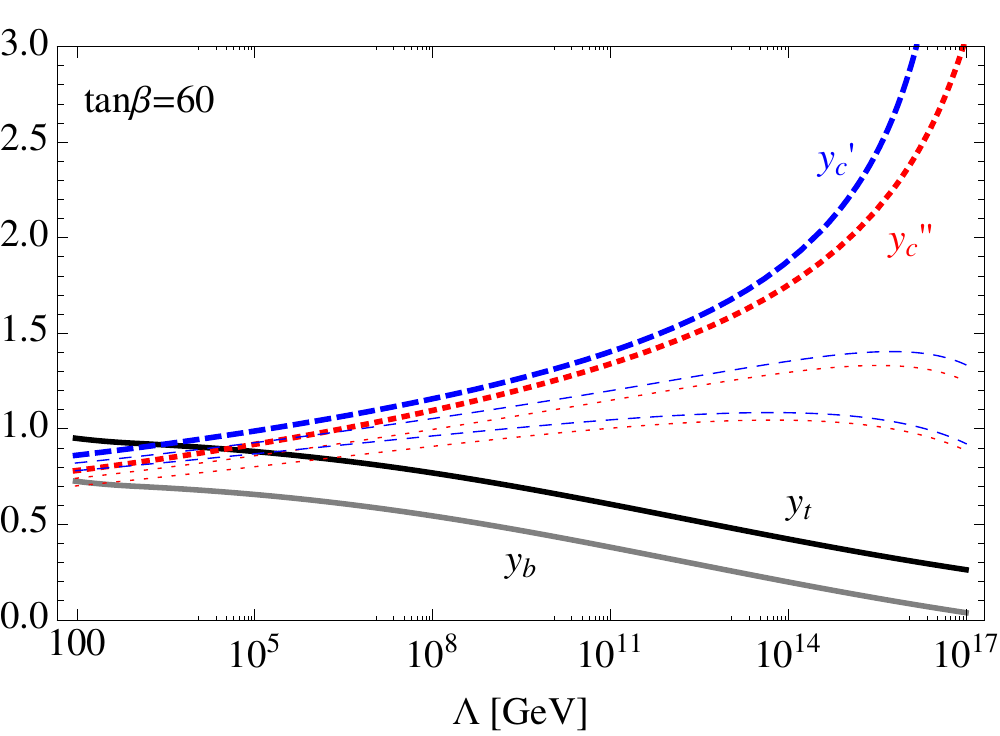}
\caption{The RGE of the Yukawa couplings for $\tan\beta=2$ (left) and $\tan\beta=60$ (right). Shown are the evolution of the new lepton Yukawa couplings $y_c'$ and $y_c''$ for various input values at the high scale, to illustrate the fixed points. }
\label{fig:rgeY}
\end{figure}

For the numerical analysis, we used the $\overline{\text{MS}}$ running masses of the top and bottom quarks at the weak scale, $M_t(M_Z) \approx 165$~GeV and $M_b(M_Z)\approx 2.7$~GeV. Then the top and bottom Yukawas are given by
\begin{align}
	y_t(M_Z) & = \frac{M_t(M_Z)}{v \sin\beta}\,, \\
	y_b(M_Z) & = \frac{M_b(M_Z)}{v \cos\beta} \frac{1}{1+ \Delta M_b}\,.
\end{align}
The SUSY-QCD correction to the bottom mass, $\Delta M_b$, is relevant in the large $\tan\beta$ regime and was included in our analysis.

The running of the couplings for small and large $\tan\beta$ is shown in Fig.~\ref{fig:rgeY}. At the weak scale, the couplings $y_c'$, $y_c''$ are largely independent of their precise magnitude at the unification scale. In particular, we note that Yukawas of order 0.5-0.9 at the weak scale are natural given order one input values at the high scale. For $\tan\beta \sim 60$ the bottom Yukawa coupling is comparable to $y_t$, and the upper bounds on the new Yukawas are $y_c' \lesssim 0.9$ and $y_c'' \lesssim 0.8$, while for small values of $\tan\beta\sim 2$ the upper bounds are approximately given by $y_c' \lesssim 0.94$ and $y_c'' \lesssim 0.72$. 

Comparing with the running of the top and bottom Yukawa coupling, we note that the $\alpha_3$ contribution tends to suppress the quark Yukawas at high scales, or, reversing the argument, leads to larger couplings at low scales. If one would like to obtain larger $y_c'$ and $y_c''$ at the weak scale, one could therefore either give up on perturbativity up to the unification scale or introduce additional interactions at intermediate scales that enhance the leptonic Yukawa couplings. 

Before moving to the next section, let us briefly comment on the evolution of the Higgs quartic couplings below the SUSY breaking scale. At the weak scale, the particle content of our model is that of the standard model with an additional Higgs doublet, a set of vector-like leptons as in~\cite{ASW}, and potentially their superpartners. We have seen in~\cite{ASW} that additional order one Yukawa couplings in conjunction with a Higgs mass of 125~GeV lead to vacuum instabilities due to the RGE of the Higgs quartic coupling, but only above the TeV scale. Supersymmetry is expected to be restored at least partially around that scale\footnote{In particular stops and weak gauginos are still allowed to be significantly lighter than one TeV.}, above which the quartic couplings in the Higgs potential are determined through supersymmetric relations and stop being a threat to the stability of the vacuum. 
\section{Higgs Properties}\label{sec:higgs}
The MSSM Higgs sector consists of the two Higgs doublets $H_u$, $H_d$ that acquire vacuum expectation values (VEV) $v_u$ and $v_d$, respectively. The VEVs are subject to the constraint $v_u^2 + v_d^2 = (174~{\rm GeV})^2$ and are therefore parametrized as $v_u = v \sin\beta$ and $v_d = v \cos\beta$. 

The physical spectrum contains two neutral CP even Higgs bosons $h^0$ and $H^0$, a neutral CP odd boson $A^0$ and a charged scalar $H^\pm$. The neutral mass eigenstates are linear combinations of the neutral doublet components $H_u^0$, $H_d^0$, with the mixing parameterized as:
\begin{align}
\begin{pmatrix} H_u^0\\H_d^0\end{pmatrix} & = \begin{pmatrix} v_u\\v_d\end{pmatrix} + \frac{1}{\sqrt{2}}\begin{pmatrix} \text{cos}\,\alpha & \text{sin}\,\alpha\\-\text{sin}\,\alpha & \text{cos}\,\alpha \end{pmatrix} \begin{pmatrix} h^0\\H^0 \end{pmatrix} + \frac{i}{\sqrt{2}}\begin{pmatrix} \text{cos}\,\beta & \text{sin}\,\beta\\-\text{sin}\,\beta & \text{cos}\,\beta \end{pmatrix} \begin{pmatrix} G^0\\A^0 \end{pmatrix}.
\end{align}
Here $\alpha$ is the CP-even mixing angle, $G^0$ is the neutral Goldstone boson that can be removed by going to unitary gauge, and 
by convention $h^0$ is defined to be the lightest of the CP-even mass eigenstates.

A particularly interesting regime is the so called decoupling limit, which is characterized by a large mass scale $m_A$ for the non-standard Higgs bosons $H^0$, $A^0$ and $H^\pm$. In this limit, the lightest CP even Higgs $h^0$ becomes SM like, and the mixing angle $\alpha$ is determined by $\alpha = \beta -\pi/2$.   Supersymmetric constraints on the potential constrain its mass to be 
\begin{align}
	m_{h^0}^2 & = m_Z^2 \cos^2(2\beta) + \text{radiative corrections.}
\end{align}
For the reminder of this paper, we will work in the decoupling limit with $m_A = 1$~TeV, and further assume that the MSSM superpartners do not significantly modify the Higgs properties. This is strongly motivated by the recent discovery of a scalar resonance with a mass of around $125$~GeV and with properties consistent with a SM Higgs boson. While this mass exceeds the lightest Higgs mass in the MSSM at the tree level, it is well known that radiative corrections can bring this mass in agreement with the observation.

The dominant $h^0$ production channel at the LHC is $gg\rightarrow h^0$, where the largest contribution comes from the top loop. As new leptons and sleptons added to the MSSM are not colored, the production channel is not affected significantly. This is consistent with the data. The new particles affect the decay widths of $h^0$ only at the loop level. Hence they can have significant effects only on the loop induced Higgs decays like $h^0\rightarrow\gamma\gamma$ and $h^0 \to Z\gamma$ which are not present at the tree-level in the MSSM.  

At the time of the new boson discovery announcement, both ATLAS and CMS experiments at the LHC reported an excess of events in the $p p \to h^0 \to \gamma\gamma$ channel with respect to the SM expectations~\cite{Chatrchyan:julyhiggs}. After including the full 2012 dataset into the analysis, the ATLAS collaboration continues to observe an increased signal strength of $1.65^{+0.34}_{-0.30}$ in the diphoton channel~\cite{ATLASmoriond}, while the signal strength measured by the CMS collaboration has decreased to $0.78^{+0.28}_{-0.26}$ or $1.11^{+0.32}_{-0.30}$, depending on the analysis method~\cite{Chatrchyan:moriond}. 
Naively averaging the ATLAS and CMS results, we obtain a signal strength of $1.14\pm 0.21$ or $1.37\pm 0.22$, depending on which of the CMS results is used.

In the following we will discuss the effects of the new leptons and sleptons on the $h^0 \to \gamma\gamma$ decay and on the mass of the Higgs. We focus on conditions to obtain a moderate enhancement of the diphoton rate, which is well in agreement with the data. 

\subsection{$h^0\rightarrow\gamma\gamma$ Width}

The matrix of the leptons (sleptons) to $h^0$ couplings is given by the gradients of the lepton (slepton) mass matrix with respect to the Higgs VEVs, projected onto the $h^0$ eigenstate: 
\begin{align}
Y_{\tilde{E}h^0}&=\sin\beta \frac{\partial \mathcal{M}^2_{\tilde{E}}}{\partial v_u}+\cos\beta \frac{\partial\mathcal{M}^2_{\tilde{E}}}{\partial v_d}\,,\\
Y_{Eh^0}&=\sin\beta\frac{\partial \mathcal{M}_{E}}{\partial v_u}+\cos\beta \frac{\partial\mathcal{M}_{E}}{\partial v_d}\,,
\end{align}
where ${\cal M}_{\tilde{E}}$ is the slepton mass matrix~(\ref{eqn:slepmass}) while ${\cal M}_E$ is any one of the off-diagonal $2\times 2$ blocks of the lepton mass matrix~(\ref{eqn:lepmass}). We also have used the decoupling relation $\alpha = \beta - \pi/2$.
Therefore, the couplings of $h^0$ to the new charged slepton and lepton mass eigenstates are given as 
\begin{align}
	C_{\tilde{E}h^0} & =   U_S^\dagger Y_{\tilde{E}h^0}U_S\,,   \\
	C_{Eh^0} &= U_L^\dagger Y_{E h^0} U_R\,,    
\end{align}
where $U_S$ is the unitary matrix which diagonalizes $\mathcal{M}_{\tilde{E}}$, while $U_L$ and $U_R$ are the unitary matrices that diagonalize $\mathcal{M}_{E}\mathcal{M}_{E}^\dagger$ and $\mathcal{M}_{E}^\dagger\mathcal{M}_{E}$, respectively.

The new charged leptons and sleptons, $E_i$ and $\tilde{E}_i$, contribute to the $h^0$ decay to two photons at the one loop level. 
In the mass basis, the contributions to the decay are given by
\begin{align}\label{eqn:hgg_loop}
	\Gamma_{h \to \gamma\gamma} \propto \left| A_1(\tau_w) + \frac{4}{3}A_{1/2}(\tau_t) + \sum_{i=1,2}\frac{C_{Eh^0_{ii}} v}{M_{E_i}}A_{1/2}(\tau_{E_i}) + \frac{1}{2}\sum_{i=1}^4\frac{C_{{\tilde{E}}h^0_{ii}}v}{M^2_{{\tilde{E}}_i}}A_0(\tau_{{\tilde{E}}_i}) \right|^2\,,
\end{align}
where the loop functions for spin 0, spin 1/2 and spin 1 particles are given by~\cite{Djouadi:2005gi}
\begin{align}
A_0(\tau) & = \frac{f\left(\tau\right)-\tau}{\tau^2} & \text{for spin  }\; 0 \,,\\
A_{1/2} (\tau) & = \frac{2\left(\tau+\left(\tau-1\right)f\left(\tau\right)\right)}{\tau^2} & \text{for spin  }\; \frac{1}{2} \,,\\
A_1(\tau) &=-2-\frac{3}{\tau}-\frac{3\left(2\tau-1\right)f\left(\tau \right) }{\tau^2 } & \text{for spin  } \;1\,,
\end{align}
with
\begin{align}
\tau_x&=\frac{m_h^2}{4m_x^2}\\
f\left(\tau\right) &= \begin{cases} \arcsin^2\sqrt{\tau}  & \text{for } \tau\leq1\,,\\
\displaystyle
-\frac{1}{4}\left(-i\pi+\log\left[\frac{1+\sqrt{1-\tau^{-1} }}{1-\sqrt{1-\tau^{-1}}}\right]\right)^2\quad \quad &\text{for } \tau>1\,,
\end{cases}
\end{align}
and $m_x$ is the mass of the particle running in the loop. 
It is instructive to consider the asymptotic values of the loop functions for $\tau_x\ll 1$, i.e. when the new particle masses are much larger than the half of the lightest Higgs boson mass. Asymptotically $A_{1/2}(\tau \to 0) = 4/3$ and $A_0(\tau \to 0) = 1/3$, while $A_{1/2}(\tau) > 4/3$ and $A_0(\tau) > 1/3$ for $0<\tau<1$. Note in particular that the SM contribution for $m_h=125$~GeV is $A_1(\tau_w)=-8.3$ from the W-boson loop and $\frac 4 3 A_{1/2}(\tau_t) = 1.8$ from the top quark loop. 

Since the new leptons don't affect the Higgs production channels, the effect on the di-photon search channel at the LHC is fully described by the ratio
\begin{align}\label{eqn:rgamma}
	R_{\gamma \gamma} =\frac{\sigma(pp \to h)}{\sigma_{\rm SM}(pp \to h)} \frac{\Gamma(h \to \gamma\gamma) }{\Gamma(h\to \gamma\gamma)_{\rm SM}} = \frac{\Gamma(h \to \gamma\gamma) }{\Gamma(h\to \gamma\gamma)_{\rm SM}}\,.
\end{align}
In the limit of  $M_L\,,M_E\,,\mu\rightarrow0$ and vanishing soft SUSY-breaking mass terms, the prefactors $C_{Eh^0_{ii}} v/M_{E_i}$ and $C_{\tilde{E}h^0_{ii}} v/M^2_{\tilde{E}_i}$ in~(\ref{eqn:hgg_loop}) go to $\tan\beta$ and $\cot\beta$, respectively, which are positive numbers. This leads to destructive interference between the dominant $W$ boson contribution and the charged lepton and slepton loops, thus $R_{\gamma\gamma}< 1$ is expected in this limit. 

In order to understand how $R_{\gamma\gamma} > 1$ can be achieved, we note that in the asymptotic limit the charged slepton contribution to the amplitude can be written as
\begin{align}\label{eqn:higgsS_low_energy}
\Delta_{\tilde{E}} & = A_{0}(0) \sum_{i} \frac{v\,C_{\tilde{E}h_{ii}}}{M_{\tilde{E}_i}^2} = \frac{1}{3} v\left(\left(\sin\beta \frac{\partial}{\partial v_u} \log\det{\cal M}^2_{\tilde{E}}\right)+\left(\cos\beta \frac{\partial}{\partial v_d} \log\det{\cal M}^2_{\tilde{E}} \right)\right)\,.
\end{align}
Similarly for leptons,
\begin{align}\label{eqn:higgs_low_energy}
\Delta_E&=A_{1/2}(0) \sum_{i} \frac{v\,C_{Eh_{ii}}}{M_{E_i}} = \frac{4}{3} v\left(\left(\sin\beta\frac{\partial}{\partial v_u} \log\det{\cal M}_E\right)+\left(\cos\beta \frac{\partial}{\partial v_d} \log\det{\cal M}_E \right)\right)\,.
\end{align}
The second equality is a consequence of the Higgs low energy theorem~\cite{Ellis:1975ap,Shifman:1979eb,Falkowski:2007hz,Carena:2012xa}, but can also be understood by noting that the trace of a matrix is basis independent and using $\log\det {\cal M} = {\rm tr}\log{\cal M}$. It is useful to evaluate the derivative partially, which leads to 
\begin{align}\label{eqn:enhancementS}
\Delta_{\tilde{E}}&= \frac{v}{3(\det{\cal M}^2_{\tilde{E}})}\left(\sin\beta \frac{\partial \det{\cal M}^2_{\tilde{E}}}{\partial v_u}+ \cos\beta \frac{\partial \det {\cal M}^2_{\tilde{E}}}{\partial v_d}\right),
\end{align}
\begin{align}\label{eqn:enhancementF}
\Delta_E&= \frac{4v}{3(\det{\cal M}_{E})}\left(\sin\beta \frac{\partial \det{\cal M}_E}{\partial v_u}+ \cos\beta \frac{\partial \det {\cal M}_E}{\partial v_d}\right).
\end{align}
To obtain an enhancement of  $R_{\gamma\gamma}$, it is clear $\Delta_{\tilde{E}}+ \Delta_E$ must be negative in order to have constructive interference with W boson loop. Furthermore it is evident that a large mass for all new particles will lead to a suppression of the effects due to the determinant in the denominator. This can be used to study the effects of sleptons and leptons separately, since either sector can be decoupled by introducing a sufficient amount of supersymmetry breaking. 

Conditions for constructive interference from vector-like leptons were studied in detail in~\cite{ASW}. A new effect that enters in the case of supersymmetry is the dependence on $\tan\beta$, since opposite chirality leptons can not couple to the same Higgs doublet. It follows that
\begin{align}\label{eqn:prefactor2}
\Delta_{E}&=\frac{4}{3}\frac{\tan\beta\,v^2y'_cy''_c}{(1+\text{tan}^2\beta)(-M_LM_E+\frac{\text{tan}\,\beta\,v^2y'_cy''_c}{1+\text{tan}^2\beta})}\,.
\end{align}
Both $\tan\beta = 0$ and $\tan\beta = \infty$ lead to a suppression of $\Delta_E$, while the effect is maximal for $\tan\beta$ of order one. Even then, the effect on $R_{\gamma\gamma}$ is reduced compared to a non-supersymmetric model where leptons of both chiralities are allowed to couple to the same Higgs. 

The slepton effects are slightly more involved. Both the vector mass terms $M_L$, $M_E$ and the $\mu$ term appear in off-diagonal elements of the mass matrix, such that in general all four charged slepton fields will mix heavily. 

Absence of tachyons implies that $\det {\cal M}_{\tilde{E}}^2$ must be positive, therefore the derivatives in~(\ref{eqn:enhancementS}) must be negative in order to obtain an enhanced $h\to\gamma\gamma$ rate. In order to see how the different terms in the mass matrices influence $\Delta_{\tilde{E}}$, it is instructive to consider the large $\tan\beta$ limit. Then all terms proportional to $v_d$ can be neglected. Further setting the soft breaking parameters to zero, one obtains
\begin{align}\label{eqn:prefactor1}
\Delta_{\tilde{E}}&=-\frac{1}{3}\frac{2 v_u^2 y'^2_c (M_E^2 M_L^2 + 2 (M_E^2 + M_L^2) v^2_u y''^2_c + 3v^4_u y''^4_c)\mu^2}{M_E^4 M_L^4 - v^2_u y'^2_c (M_E^2 + v^2_u y''^2_c) (M_L^2 + v^2_u y''^2_c) \mu^2} \\
&\quad \to 
 -\frac{1}{3}\frac{2 v_u^2 y'^2_c \mu^2}{M_E^2 M_L^2 - v^2_u y'^2_c   \mu^2}\,.
\end{align}
In the last line we have further set $y_c'' \to 0$. 
This result highlights the importance of $\mu^2$ for obtaining a large contribution to $R_{\gamma\gamma}$. While the numerator grows linearly with $\mu^2$, the denominator can be held roughly constant by appropriately adjusting $M_E$ and $M_L$, such that the slepton masses are in agreement with the LEP limits. Very large values for $\Delta_{\tilde{E}}$ can therefore be obtained at the expense of some tuning between the mass parameters. 

In section~\ref{sec:vacuum} we will see that vacuum stability considerations will lead to an upper bound on $\mu^2$ in the presence of large Yukawa couplings $y_c'$ and $y_c''$ and small slepton masses, which will limit the amount by which $R_{\gamma\gamma}$ can be enhanced. Numerical results that cover the full parameter region of our model and show the maximal possible enhancement are shown in section~\ref{sec:results}. 

\subsection{One Loop Corrections to the $h^0$ mass}
To compute the one loop corrections to the $h^0$ mass, we use the one loop effective potential approximation. A supermultiplet $i$ with scalar of mass $M_{S_i}$ and fermion of mass $M_{F_i}$ contributes 
\begin{align}
\Delta V_i&=\frac{N_c}{32\pi^2}\left[M_{S_i}^4\left(\log \left(\frac{M_{S_i}^2}{Q^2}\right)-\frac{3}{2}\right)-M_{F_i}^4\left(\log\left(\frac{M_{F_i}^2}{Q^2}\right)-\frac{3}{2}\right)\right]
\end{align}
to the one-loop effective potential. Here $Q$ is the renormalization scale.  Then, as given in \cite{Martin:2009bg}, the correction to the tree-level $m_{h^0}^2$ is 
\begin{align}\label{eqn:masscorr}
\Delta m_{h^0}^2 &=\left(\frac{\text{sin}^2\beta}{2}\left[\frac{\partial^2}{\partial v_u^2}-\frac{1}{v_u}\frac{\partial}{\partial v_u}\right]+\frac{\text{cos}^2\beta}{2}\left[\frac{\partial^2}{\partial v_d^2}-\frac{1}{v_d}\frac{\partial}{\partial v_d}\right]+\text{sin}\,\beta\,\text{cos}\,\beta\frac{\partial^2}{\partial v_u\partial v_d}\right)\sum_i\Delta V_i \,.
\end{align}
A notable modification of $R_{\gamma\gamma}$ requires at least some of the new leptons or sleptons to be light. In the absence of soft breaking terms, only $\mu$ will induce SUSY breaking in the new lepton sector, and therefore corrections to the Higgs mass will remain small. Larger corrections are possible if soft terms induce a sizable mass splitting between the sleptons and leptons, and it is worth asking whether these corrections can improve upon the fine tuning in the MSSM. 
\begin{figure}
\center
\includegraphics[width=.48\textwidth]{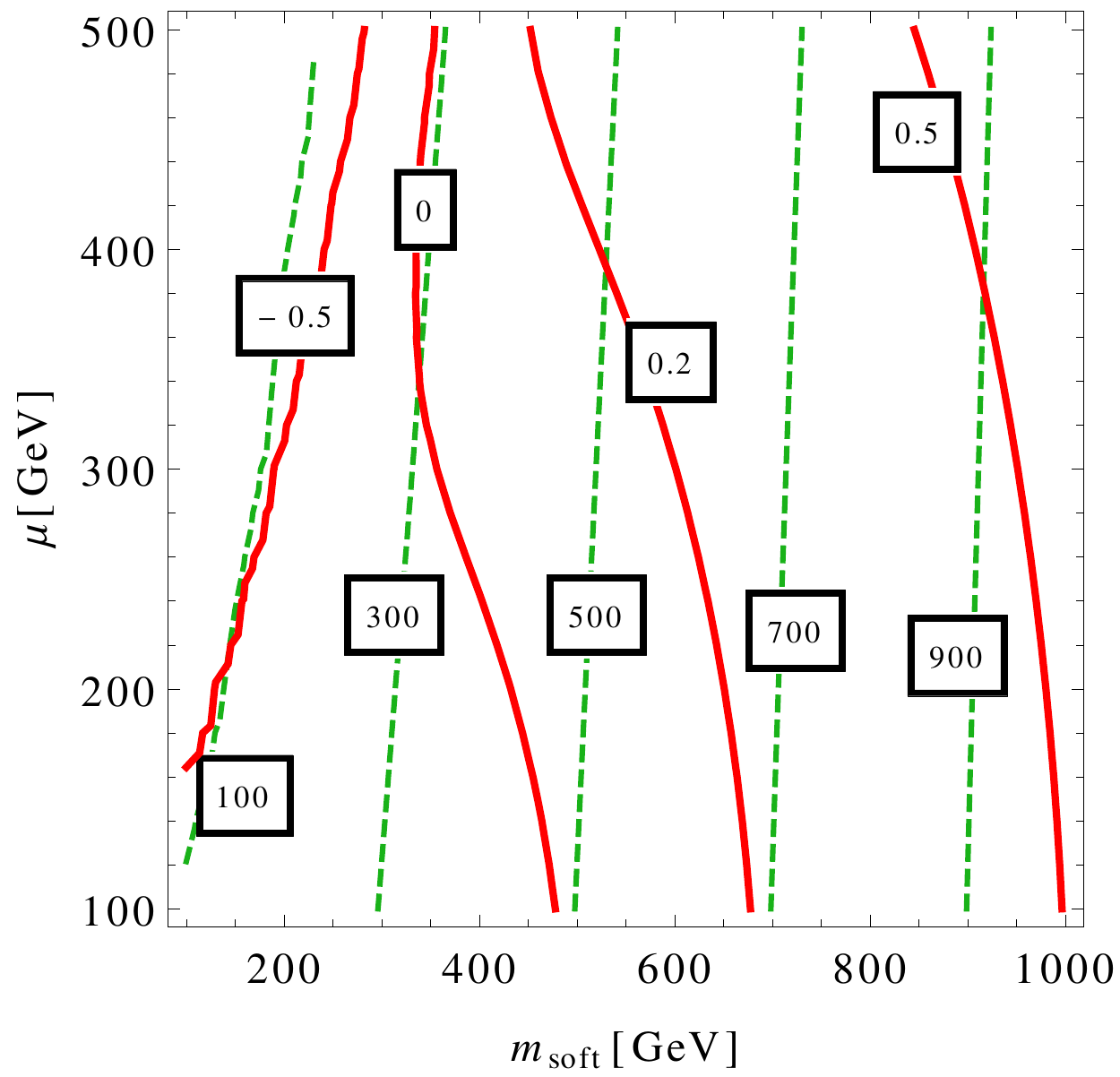}\hspace*{.5cm}
\includegraphics[width=.48\textwidth]{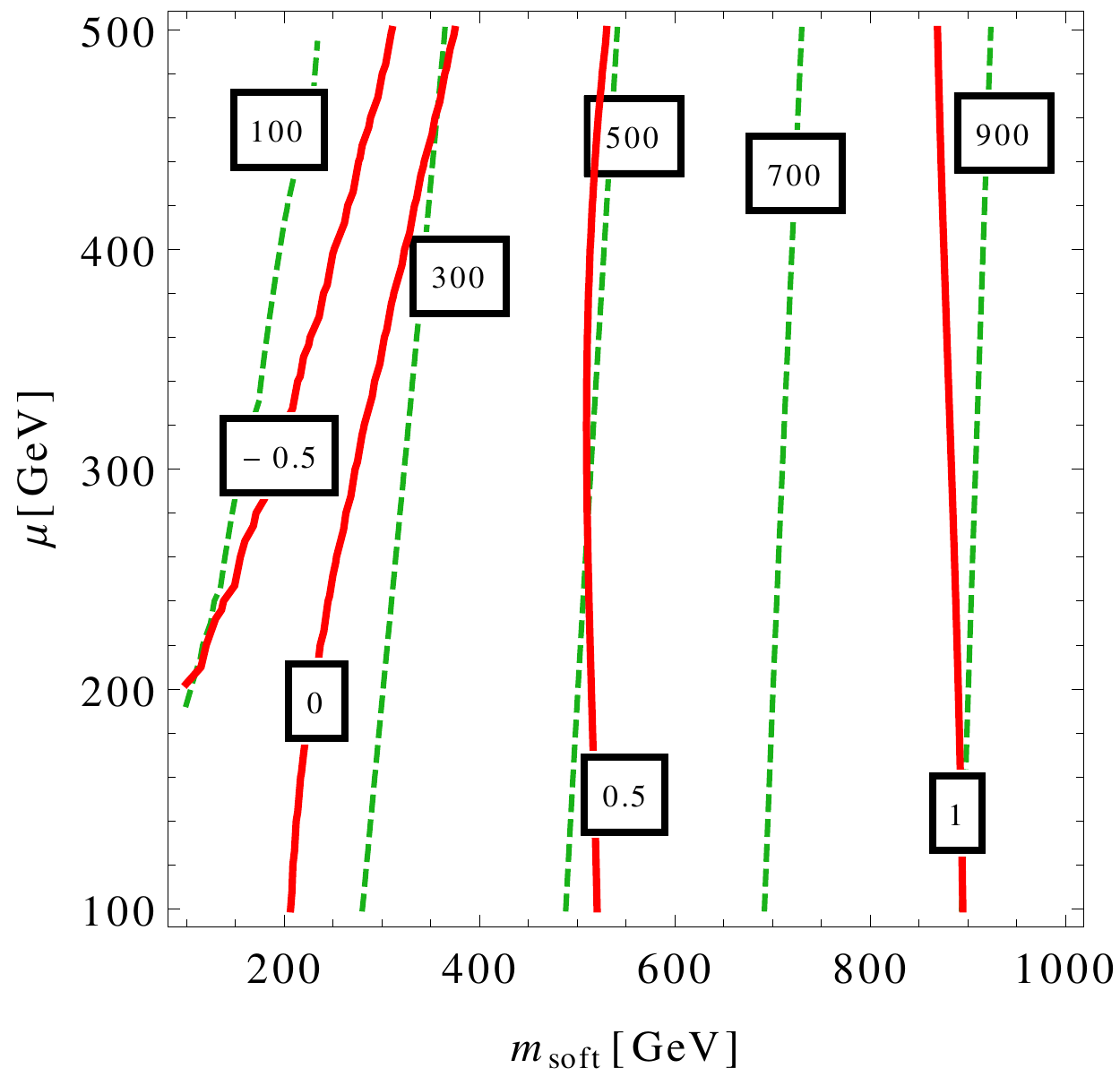}
\caption{Contributions to the Higgs mass $\Delta m_{h^0}$ in the $m_{\rm soft} - \mu$ plane for $\tan\beta=2$ (left) and $\tan\beta=60$ (right). The red (solid) curves are contours of constant $\Delta m_{h^0}$ while the green (dashed) contours show the mass of the lightest slepton state. The remaining parameters are $M_L = M_E = 200$~GeV, $y_c' = 0.9$, $y_c'' =0.7$, and we assume equal soft masses $m_i^2 = m_{\rm soft}^2$ for all sleptons. 
}
\label{fig:higgsmass}
\end{figure}

Compared to the corrections typically obtained from top-stop splitting, we can expect more moderate contributions due to the absence of a color factor and because the Yukawa couplings, that enter in the fourth order in the Higgs mass, are smaller than the top Yukawa coupling. 
In Fig.~\ref{fig:higgsmass} we show contours of $\Delta m_{h^0}$ in the $m_{\rm soft}-\mu$ plane. For these plots, we have assumed a base Higgs mass of $m_0=120$~GeV in the MSSM and we show the correction $\Delta m_{h^0} = \sqrt{m_0^2 + \Delta m_{h^0}^2} - m_0$, with $\Delta m_{h^0}^2$ coming from~(\ref{eqn:masscorr}). The sleptons are taken to have equal soft masses $m_i^2 = m_{\rm soft}^2$, while the soft breaking $a$ and $b$ terms are set to zero. The fermionic mass scale is set to $M_L = M_E = 200$~GeV and we use $y_c' = 0.9$ and $y_c''=0.7$, close to their fixed point values. 

It is easy to see that even for TeV scale slepton masses, the Higgs mass is lifted by at most one GeV, while the corrections are negligible when both leptons and sleptons are close to the weak scale. It is possible to obtain larger corrections by adding large $a$ terms to the SUSY breaking Lagrangian. However these soft terms can potentially destabilize the vacuum beyond what will be discussed in the next section, so we refrain from introducing them. 

Clearly the new lepton sector can not alleviate the fine tuning problem in the MSSM, and furthermore will not be sufficient to lift the Higgs mass to $125$~GeV in the low $\tan\beta$ case. Instead the necessary contributions could arise from SUSY breaking in the vector-like quark sector that is not being discussed in this paper (see e.g.~\cite{Ajaib:2012eb} for a recent discussion).

\section{Vacuum Stability}\label{sec:vacuum}
To analyze the vacuum structure of our model, we need the full scalar potential for the Higgs scalars and the new sleptons. We assume that all other scalar fields have masses large enough to avoid any stability problems. The scalar potential is obtained from the superpotential as 
\begin{align}\label{eqn:potential}
	V & = \sum_{\phi_i} \left| \frac{\partial W }{\partial \phi_i} \right|^2 + \frac{1}{2}\sum_{a}g_a^2(\sum_{\phi_i} \phi_i^*T^a\phi_i)^2+ V_{\rm soft} + \text{radiative corrections}\,. 
\end{align}
where the sum runs over $\phi_i \in \left\{ h_u^0, h_d^0, \tilde{e}_L',\bar{\tilde{e}}_R',\tilde{e}_L'',\bar{\tilde{e}}_R''\right\}$, while fields that are not in danger of acquiring a VEV, like the MSSM superpartners or the sneutrinos, are omitted.  
For the different contributions to the potential, we obtain
\begin{align}
	V_F  =& \,\left| y_c' \tilde{e}_L' \bar{\tilde{e}}_R' - \mu h_u^0 \right|^2 + \left| y_c'' \bar{\tilde{e}}_R'' \tilde{e}_L'' - \mu h_d^0 \right|^2 + \left| M_L \bar{\tilde{e}}_R'' + y_c' h_d^0 \bar{\tilde{e}}_R' \right|^2  \label{eqn:Fterm} \\
	& + \left| M_E\tilde{e}_L'' + y'_c \tilde{e}_L' h_d^0 \right|^2 + \left| M_L \tilde{e}'_L + y_c'' h_u^0 \tilde{e}_L'' \right|^2 + 
	\left| M_E \bar{\tilde{e}}_R'	 + y_c'' \bar{\tilde{e}}_R'' h_u^0 \right|^2,
	\notag\\
	V_D  =& \,\frac{1}{8} g_2^2 \left( (h_u^0)^2 - (h_d^0)^2 +\tilde{e}'^2_L - \bar{\tilde{e}}''^2_R  \right)^2 + \frac{1}{8} g_1^2 \left((h_u^0)^2 - (h_d^0)^2 -\tilde{e}'^2_L + \bar{\tilde{e}}''^2_R + 2 \bar{\tilde{e}}'^2_R - 2\tilde{e}''^2_L \right)^2,\label{eqn:potentialD} \\
	V_{\rm rad}  =&\, \frac{1}{8} (g_1^2 + g_2^2) \delta_H  (h_u^0)^4, \\
	V_{\rm soft}  =&\, m_{H_u}^2 (h_u^0)^2 + m_{H_d}^2 (h_d^0)^2 + B h_u^0 h_d^0+ V_{\rm soft,\ell}\,.\label{eqn:potential1}
\end{align}
For simplicity we have assumed that all parameters and all fields are real. The Higgs soft mass parameters $m_{H_u}^2$ and $m_{H_d}^2$ are usually replaced by the parameters $v$ and $\tan\beta$ that characterize the electroweak symmetry breaking minimum, 
while $B$ is related to the CP-odd Higgs mass through $m_A^2 = 2 B/\sin(2\beta)$ at the tree level. 
For large $\tan\beta$, the correction $\delta_H$ must be roughly $\delta_H \sim 1$ in order to obtain the correct Higgs mass for the SM-like Higgs in the decoupling limit, while for $\tan\beta \sim 2$ a value of $\delta_H \sim 2.5$ is necessary. 

As long as the slepton mass matrix has only positive eigenvalues, the above potential will have a local minimum characterized by $v=174$~GeV and $\tan\beta$, with a value for the potential $V(v,\tan\beta) \equiv V_0$. Additional minima with non-zero VEVs for some of the charged slepton fields can be induced by the trilinear terms in~(\ref{eqn:Fterm})~\cite{Rattazzi}. 

When the sleptons are heavier than the electroweak scale $v=174$~GeV, these additional minima typically have a potential energy larger than the electroweak minimum, and therefore do not pose a problem. However when the slepton masses are of order $v$ or below, some of the charge breaking minima might be lower than the electroweak minimum. Since relatively light sleptons are required to obtain a large enhancement of $R_{\gamma\gamma}$, it is clear that the vacuum structure of the model must be analyzed carefully.  

Let us first consider the absolute stability condition, namely the condition that $(v, \tan\beta)$ is the global minimum of the potential. In order to see the effects of the various parameters on the vacuum stability, it is instructive to derive an analytical result in the limit of $\tan\beta \to \infty$ ($v_d \to 0$) and $\tilde{e}''^2_L,\,\bar{\tilde{e}}''^2_R\rightarrow0$, corresponding to a scenario where large soft mass terms for the mirror sleptons prevent them from acquiring a VEV. Further neglecting soft $a$ and $b$ terms, the following term is added to the MSSM Higgs potential:
\begin{align}\label{eqn:abscond}
V'&=(y_c' \tilde{e}_L' \bar{\tilde{e}}_R')^2-2\mu h_u^0y_c' \tilde{e}_L' \bar{\tilde{e}}_R'+M^2_L\tilde{e}'^2_L+M^2_E\bar{\tilde{e}}_R'^2+\text{D-terms}\,.
\end{align}
\begin{figure}
\center
\includegraphics[width=.45\textwidth]{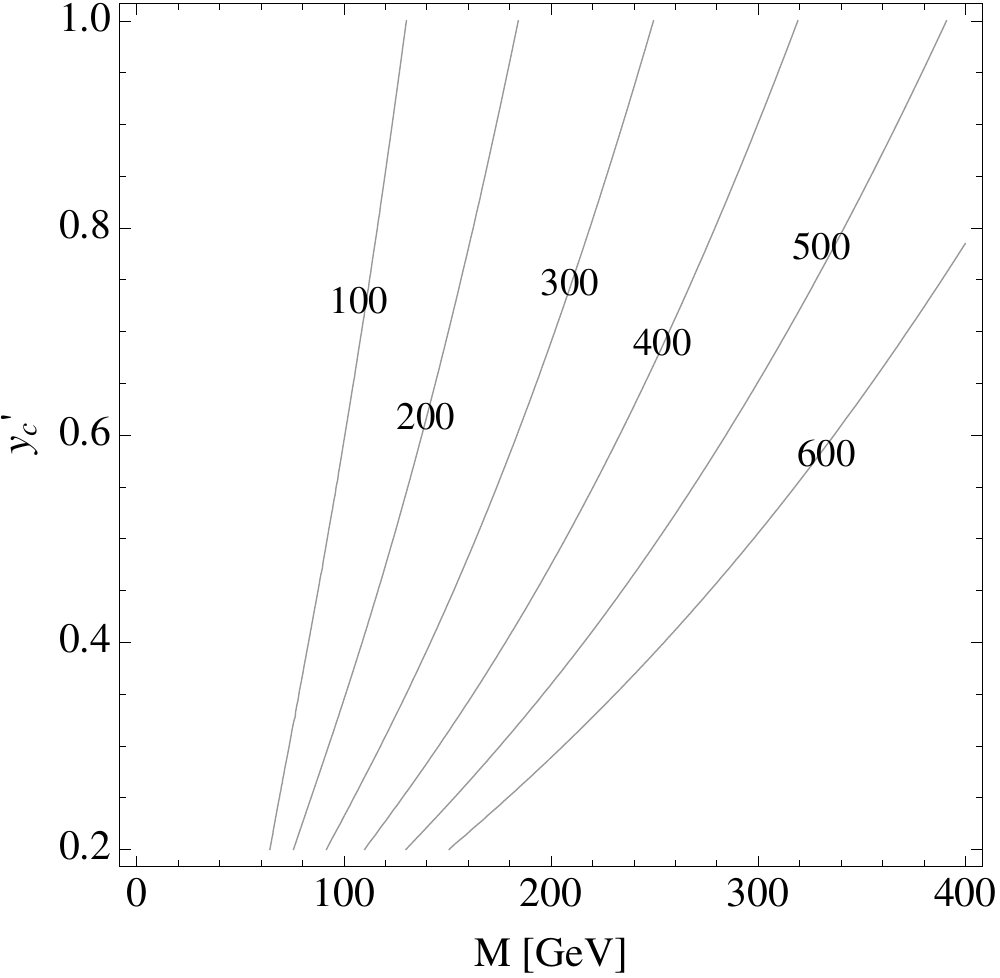}\hspace*{.5cm}
\includegraphics[width=.45\textwidth]{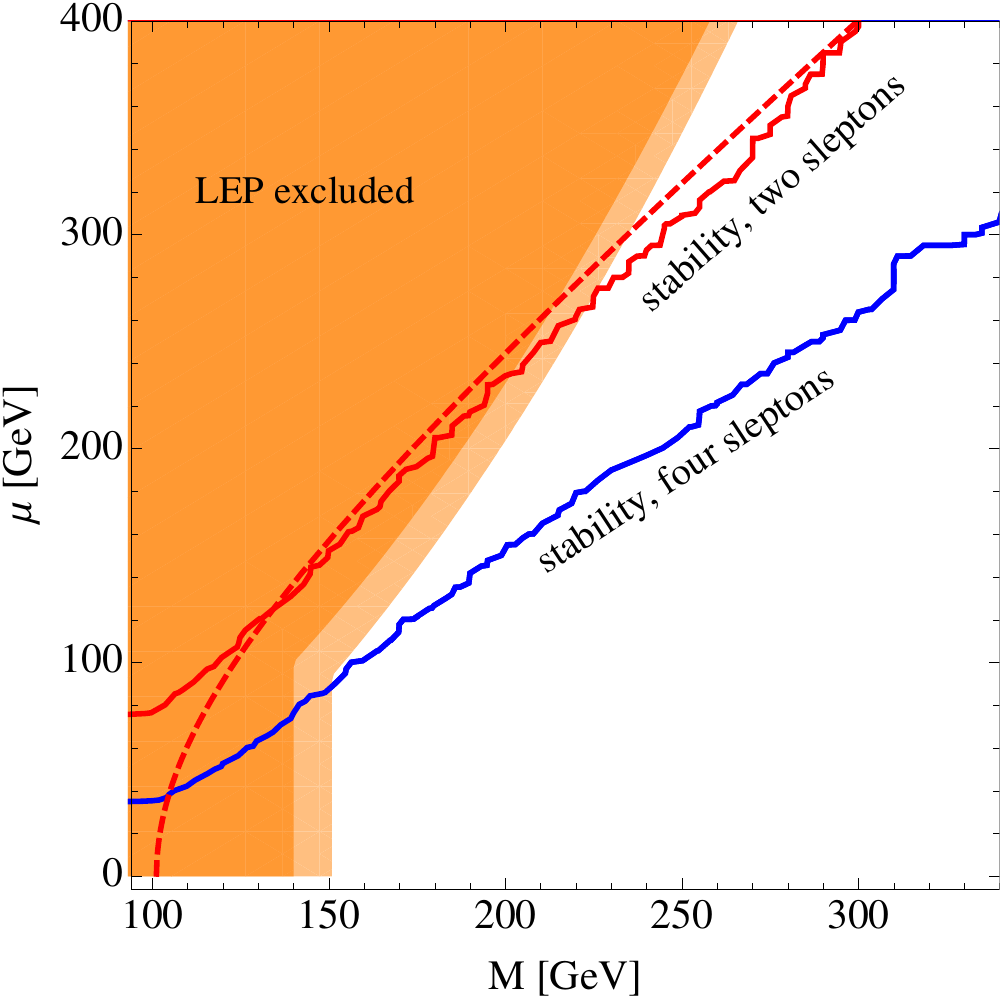}
\caption{Vacuum stability constraints on the $\mu$ parameter. Left: Allowed values for $\mu$ in the $M-y_c'$ plane in the two slepton approximation~(\ref{eqn:mubound}). 
Right: Constraints in the $\mu-M$ plane for $y_c'=0.9, y_c''=0.7$ and $\tan\beta=60$. Regions below the solid blue (red) lines are consistent with absolute vacuum stability with four (two) light sleptons. The dashed red line indicates the approximation (\ref{eqn:mubound}), while the orange region is excluded by the LEP limits on lepton and slepton masses. 
}
\label{fig:analyticResult}
\end{figure}
The instability is induced by the trilinear term proportional to $y_c'\mu$, whereas all other terms are strictly positive. A necessary (but not sufficient) condition for a deeper charge breaking minimum is $V'<0$. Therefore $V'>0$ is sufficient to guarantee stability of the Higgs potential in this limit. After some manipulations, we find
\begin{align}\label{eqn:mubound}
y'^2_c\mu^2&<\left(\frac{k}{2}+\sqrt{(1+\delta_H)\,k\left(y'^2_c+\frac{k}{4}\right)}\right)\left(M_L^2+M_E^2-v^2\sqrt{(1+\delta_H)\,k\left(y'^2_c+\frac{k}{4}\right)}\right)\,,
\end{align} 
where $k=\frac{g^2+g^2_Y}{2}$. If the Yukawa coupling is close to the RGE fixed point, this bound constrains $\mu$ to be at most of the same order as the vector masses. Allowed values for $\mu$ as a function of the Yukawa coupling $y_c'$ and of the vector mass scale $M= M_L = M_E$ are shown in Fig.~\ref{fig:analyticResult} (left). 
An sizable enhancement of the di-photon rate from the fermion sector requires $y_c'\sim 0.9$ and $M \sim 200$~GeV, which roughly translates to $\mu \lesssim 250$~GeV. 

\

When all slepton fields are included, analytic bounds on the model parameters become impossible to derive. The absolute stability condition can however easily be implemented using numerical minimization routines. An additional constraint on $\mu$ comes from the LEP limit on charged sleptons. 

The right panel of Fig.~\ref{fig:analyticResult} shows the different constraints in the $\mu-M$ plane, for $\tan\beta=60$, $y_c'=0.9$, $y_c''=0.7$ and $M_L = M_E = M$.  The lighter (darker) orange region is excluded by the conservative (optimistic) LEP limits. Regions below the solid red line are consistent with absolute stability when only two of the sleptons are light, and we see that the analytic approximation (dashed) agrees well with the numerical solution. When all four slepton fields are light, the stability constraint on $\mu$ becomes stronger, as indicated by the solid blue line. Overall it is clear that both the LEP limits and the stability constraints have to be taken into account when values of $\mu\sim M$ are being considered.

\

Since the stability limit on $\mu$ is more constraining than the LEP bound in many cases, it is worth noting that absolute stability is not a necessary condition for the model to be viable phenomenologically. After all, the measured Higgs mass of $125$~GeV implies that the standard model itself is only meta-stable~\cite{Degrassi:2012ry,Bezrukov:2012sa}. This means that while the electroweak minimum is not the global minimum of the radiatively corrected Higgs potential, the tunneling rate to the true global minimum is suppressed enough to guarantee a lifetime larger than the age of the universe.

Imposing the weaker meta-stability bound on our model could allow larger values of $\mu$ which in turn will lead to higher attainable diphoton rates, as was discussed in Sec.~\ref{sec:higgs}. If the electroweak minimum is not the global minimum of the potential, the probability to transition into the true charge breaking vacuum per unit volume and time depends on the decay rate~\cite{Coleman:1977py}
\begin{align}
\frac{\Gamma}{V}&=Ae^{-S_E}\;.
\end{align}
Here $A$ is a dimensionful parameter that is expected to be of fourth order in the electroweak scale, $A\sim v^4$, and $S_E$ is the Euclidean action of the bounce solution corresponding to the transition from Higgs vacuum to the charge symmetry breaking true vacuum. The age of the universe is about $1.37\times10^{10}$ years, which implies that $S_E\gtrsim 400$ is a necessary condition for our existence. 

The metastability bound computation is performed using the package CosmoTransitions~\cite{cosmotrans}. It finds the bounce solution for the transition from the false vacuum to the real vacuum of a multi-dimensional scalar potential by the method of path deformation. The results of this computation and its impact on $h\rightarrow\gamma\gamma$ enhancement are presented in Sec.~\ref{sec:results}.

\section{Results}\label{sec:results}
The model has several distinct regions of parameter space that can lead to an enhanced $R_{\gamma\gamma}$. In the following we will discuss the most interesting scenarios and their strengths and weaknesses. 
\subsection{Vector-like leptons}\label{sec:veclep}
First we will consider the scenario where only the new leptons are light, while the sleptons are lifted to the TeV scale by the diagonal soft mass terms in~(\ref{eqn:slepmass}). TeV scale superpartners are sufficient to protect the Higgs quartic coupling from running to negative values below the scale where supersymmetry is restored. Furthermore large mass terms for the sleptons protect this scenario from vacuum instabilities and ensure that the electroweak symmetry breaking vacuum is a global minimum. 

The conditions for obtaining a non-negligible, positive contribution to $R_{\gamma\gamma}$ are summarized in {Eq.~(\ref{eqn:prefactor2}).} First, we require that the combination $M_L M_E - v^2 y_c' y_c'' \sin\beta\cos\beta$ is positive but not too large, in order to obtain the correct sign for $\Delta_E$. Furthermore $\tan\beta$ must be of order one, otherwise the contribution will be suppressed by $1/\tan\beta$. The latter condition emerges because the effective Yukawa couplings of both leptons and mirror leptons to the lightest CP even Higgs boson are rescaled by $\sin\beta$ and $\cos\beta$ respectively, so their ratio should not be too small. 

One immediate concern is that a mass of 125~GeV for the lightest Higgs boson is difficult to obtain with $\tan\beta$ of order one within the MSSM. While we have seen in Sec.~\ref{sec:higgs} that the new slepton sector does not significantly improve the situation, there are other ways to lift the Higgs mass without affecting its low energy phenomenology. The most straightforward solution is to assume that additional one loop contributions beyond those of the top quark come from the vector-like quark sector that accompanies the leptons if they are implemented in complete SO(10) multiplets. Other possibilities include scalar singlet extensions of the MSSM or additional gauge interactions~\cite{Huo:2012tw}. For the remainder of this section, we will assume that one of these mechanisms is at work, but does not otherwise affect the phenomenology of the lightest Higgs boson.

\begin{figure}
\center
\includegraphics[width=7cm]{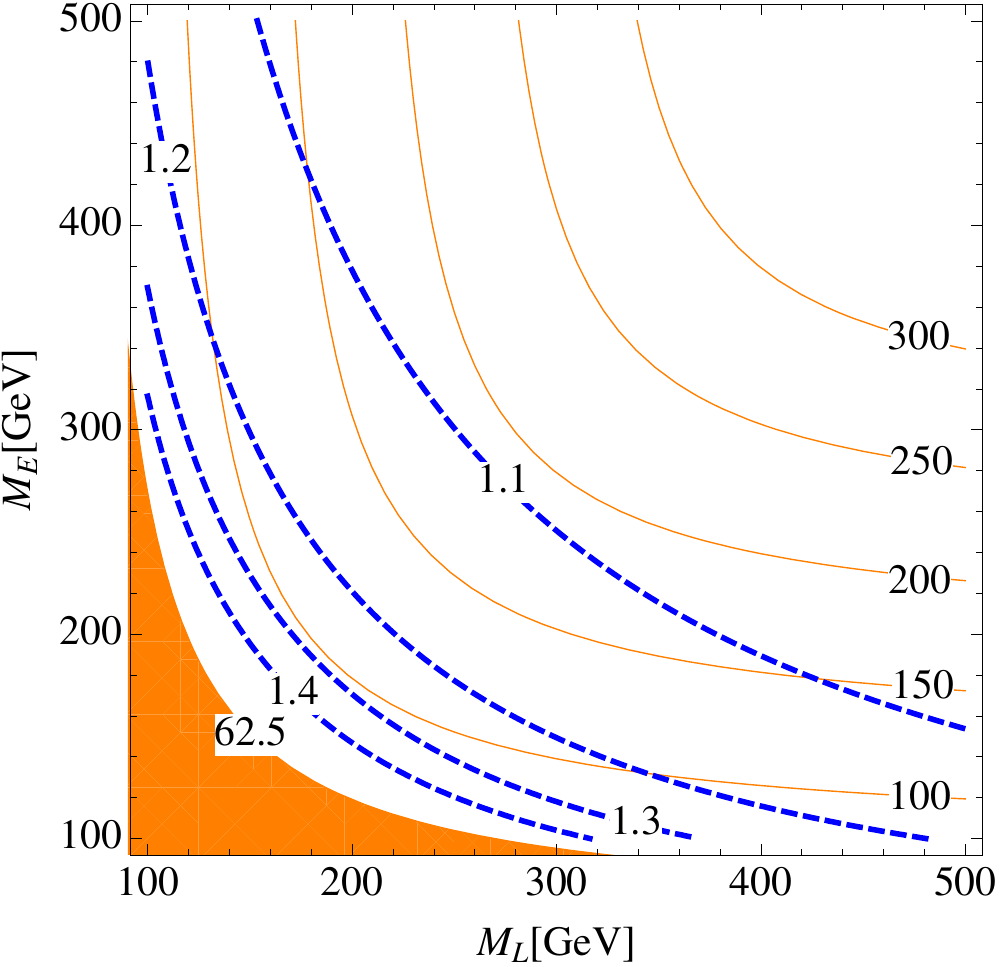}\hspace*{.4cm}
\includegraphics[width=7cm]{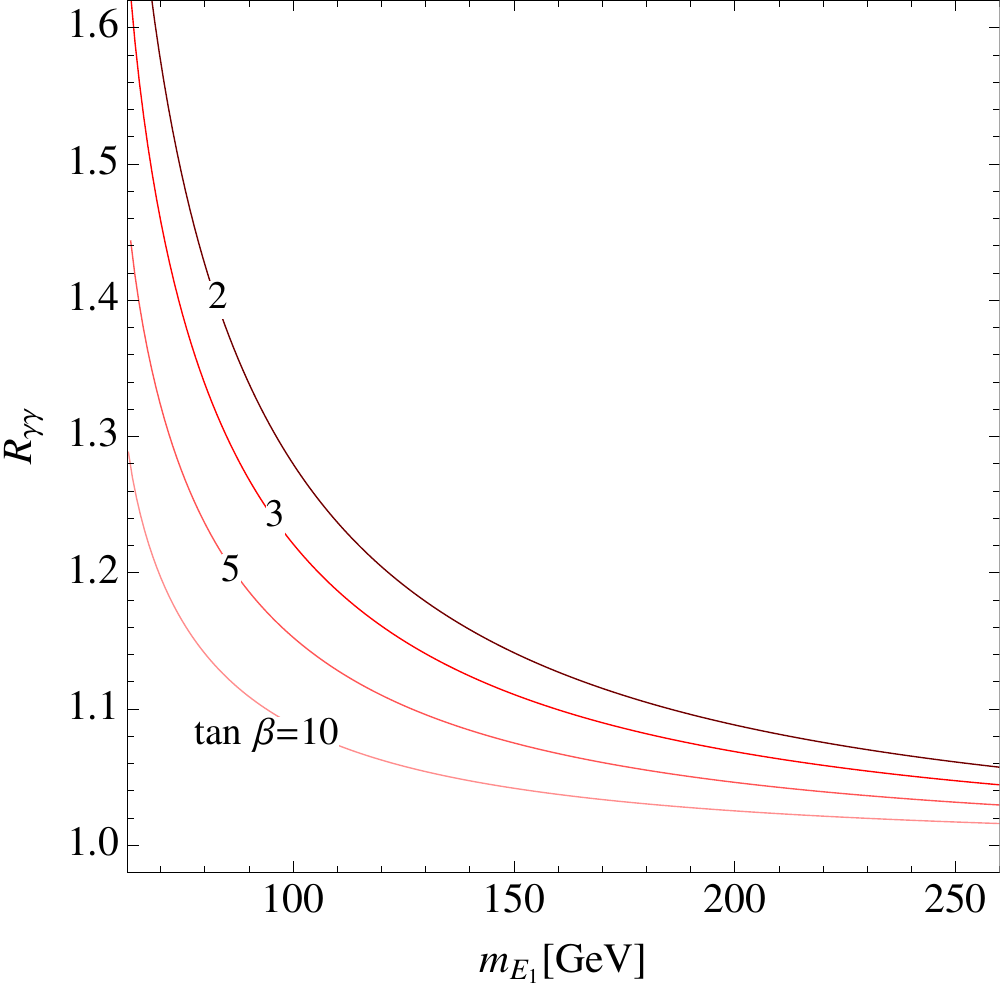}
\caption{Enhancement of the $h\to\gamma\gamma$ rate from vector leptons only, with charged Yukawa couplings {$y'_c = 0.9$ and $y''_c=0.7$.} Left: Contours of $R_{\gamma\gamma}$ in the $M_L-M_E$ plane (blue, dashed), for $\tan\beta=2$. Also shown are contours for the mass of the lightest charged lepton state (orange, solid). Right: Enhancement $R_{\gamma\gamma}$ as function of the lightest charged lepton mass $m_{E_1}$ for different values of $\tan\beta$. 
}
\label{fig:vl1}
\end{figure}

Compared with the results of~\cite{ASW}, we expect that the enhancement of $R_{\gamma\gamma}$ is suppressed roughly by $1/\tan\beta$. Note that $\tan\beta \gtrsim 1.5$ is required to ensure perturbativity of the top Yukawa coupling in the presence of sizable Yukawa couplings for the new leptons. In Fig.~\ref{fig:vl1} left we show the enhancement that can be obtained for $\tan\beta=2$ and $y_c'=0.9$, $y_c''=0.7$, close to their respective fixed point values, in the $M_L-M_E$ plane. While the features of the plot are similar, the maximal enhancement that can be obtained for a lightest lepton mass of order 100~GeV is $R_{\gamma\gamma} \lesssim 1.3$, compared with $R_{\gamma\gamma} \lesssim 1.6$ in the non-supersymmetric case. 

To further illustrate the importance of $\tan\beta$, in the right panel of Fig.~\ref{fig:vl1}, we show the contours of $R_{\gamma\gamma}$ in the plane of $\tan\beta$ and $m_{E_1}$, the mass of the lightest new lepton. For masses around 100~GeV the enhancement is reduced from around 30\% at $\tan\beta=2$ to below 10\% for $\tan\beta=10$. 

Overall one can see that the cost of imposing GUT scale stability cuts the enhancement of $R_{\gamma\gamma}$ in half. If these new fermions are the only particles that have significants effects on the Higgs phenomenology, only a modest enhancement of $R_{\gamma\gamma}$ of around 30\% is compatible with vacuum stability, perturbativity, and grand unification. 

\subsection{Supersymmetric leptons}
Here, we will first assume that at the tree level the only source of  SUSY breaking in the new lepton sector is through the $\mu$ term, such that a minimal number of parameters are added to the MSSM. When $\mu=0$ each lepton is accompanied by two sleptons with the same mass. These degenerate states are split when $\mu$ is nonzero. Therefore, in the absence of soft breaking parameters, the mass of the lightest slepton is always lower than the corresponding lepton mass. 
\begin{figure}
\center
\includegraphics[width=.47\textwidth]{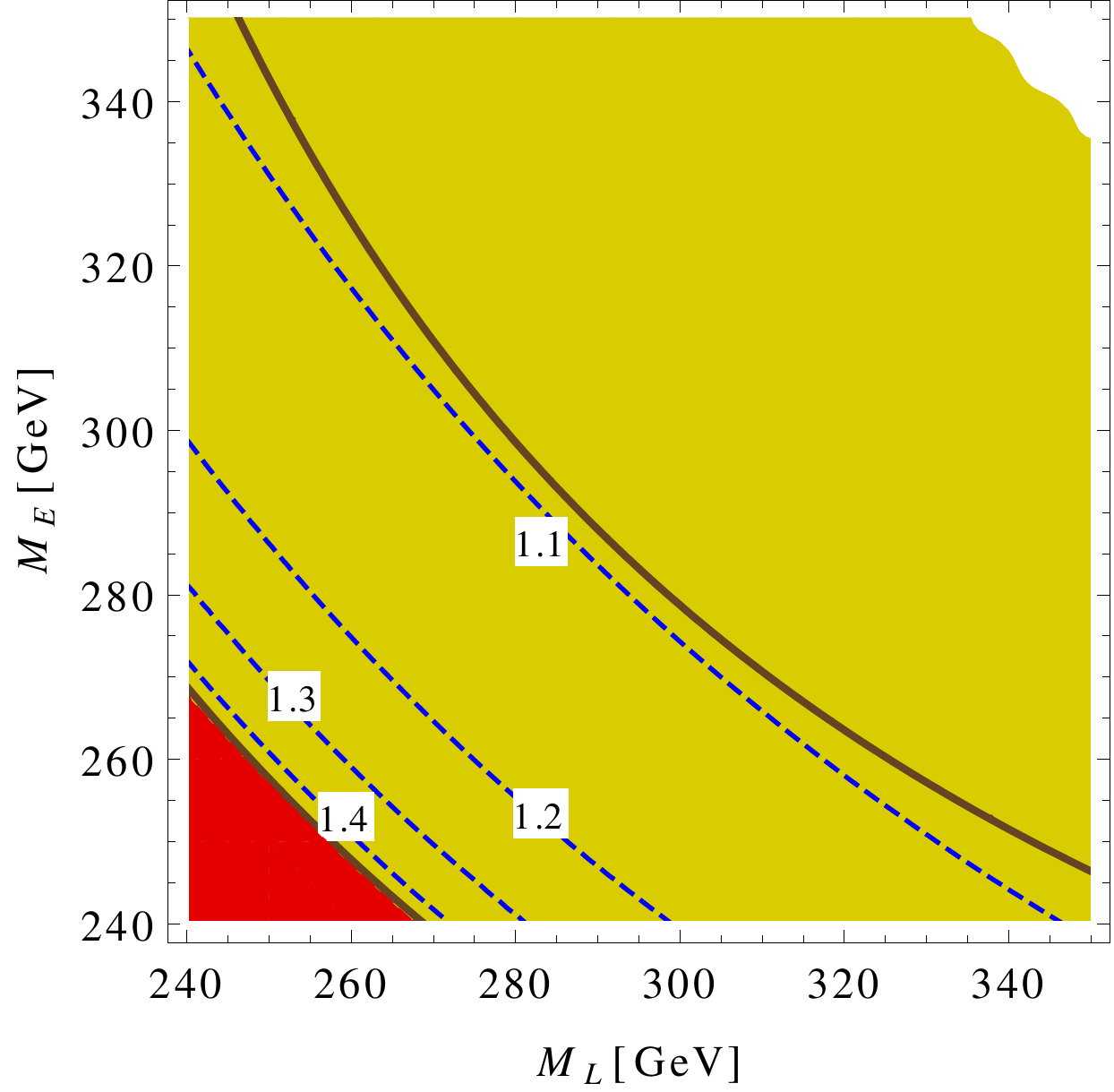}
\includegraphics[width=.47\textwidth]{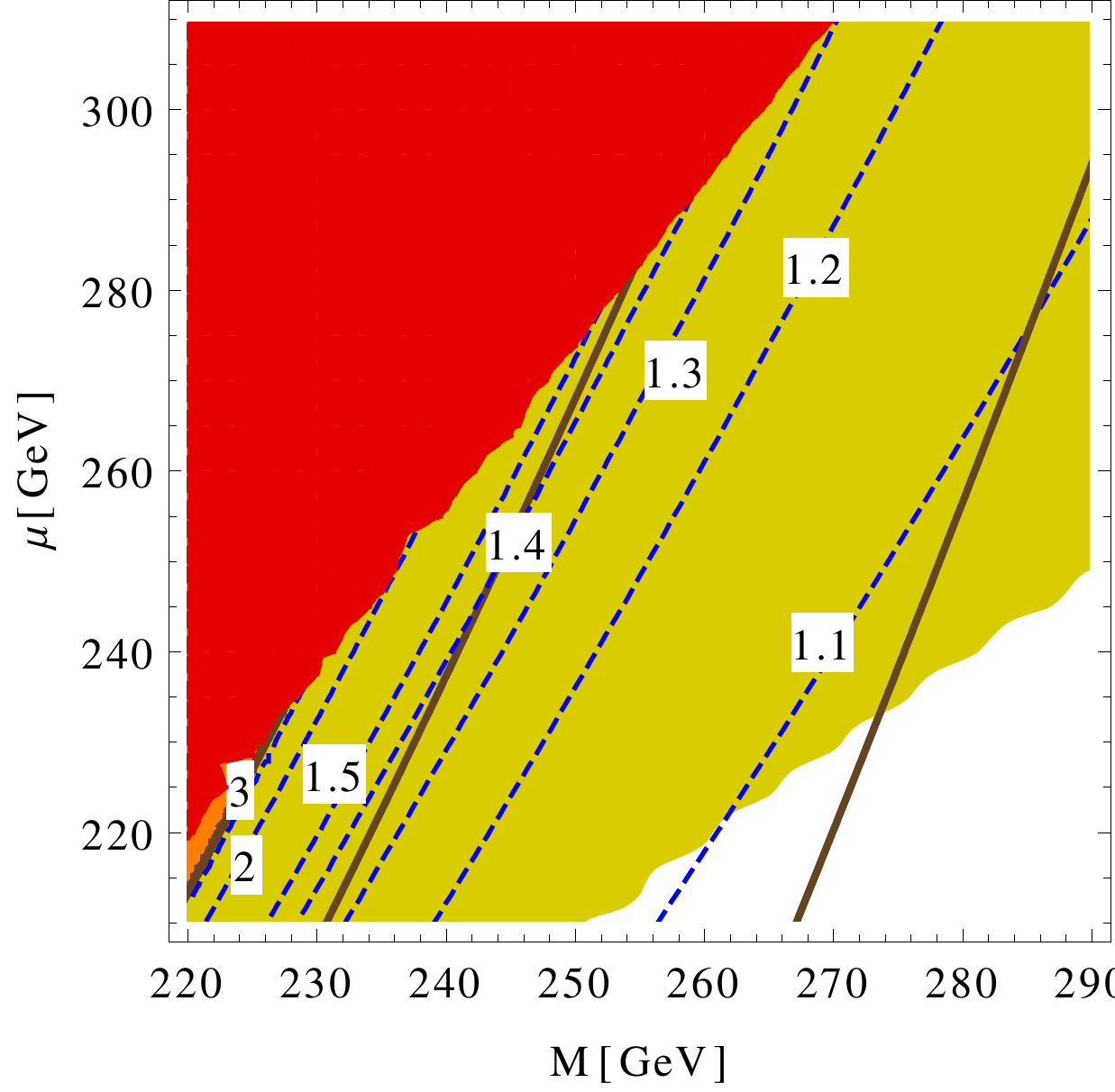}\hspace*{.5cm}
\caption{Diphoton enhancement rate with light leptons and sleptons, for $\tan\beta=60$. Regions with stable, meta-stable or unstable vacuum are shaded white, yellow and red, respectively. Regions excluded by LEP are shaded orange. 
The blue (dashed) lines are contours of constant $R_{\gamma\gamma}$. The brown (solid) lines are the slepton mass contours of 90, 150 GeV for the left and 62.5, 90, 150 GeV for the right panel reading away from the unstable region. The left panel uses $\mu=280$~GeV. }
\label{fig:hightan}
\end{figure}

Let us first consider the large $\tan\beta$ limit, where only the sleptons contribute significantly to $R_{\gamma\gamma}$. As before, the Yukawa couplings are taken to be $y_c'=0.9$ and $y_c''=0.7$, and we will usually take $M_L = M_E = M$. Since all sleptons are now light, in addition to the LEP bounds we have to impose vacuum (meta-) stability for the model to be viable. Fig.~\ref{fig:hightan} shows the diphoton rate enhancement that can be obtained for $\tan\beta=60$. Stable, meta-stable and unstable regions of parameter space are indicated in white, yellow, and red, respectively. The orange shaded region is excluded since we required $m_{\tilde{E}_1} > m_h/2$. 
Contours of constant $R_{\gamma\gamma}$ are shown (blue, dashed) as well as contours of constant $m_{\tilde{E}_1}$ (brown, solid). The left plot, where $\mu=280$~GeV, demonstrates the usual result that $M_L=M_E$ maximizes the enhancement, and we can see that an enhancement of $R_{\gamma\gamma}$ by 40\% is possible for $m_{\tilde{E}_1} \gtrsim 90$~GeV and meta-stability. It is evident from the structure of the lepton mass matrix given in Eq.~(\ref{eqn:lepmass}) that the lepton masses are well above the conservative LEP bound of $100\,\text{GeV}$ for the region of the parameter space shown here. 

The right plot of Fig.~\ref{fig:hightan} shows the interplay of the different constraints as $\mu$ is increased. Initially, the LEP bound is more constraining, and enhancements of at most 40\% are possible with $m_{\tilde{E}_1}>90$~GeV. Following the $m_{\tilde{E}_1}=90$~GeV contour to higher values of $\mu$ we note that $R_{\gamma\gamma}$ increases up to 1.5 before we eventually hit the stability bound. Lowering the lightest slepton mass is only possible for $\mu\lesssim 280$~GeV. In that region, enhancements of 100\% or more seem possible when $m_{\tilde{E}_1}$ is very close to the absolute lower bound of 62.5~GeV.

\begin{figure}
\center
\includegraphics[width=.47\textwidth]{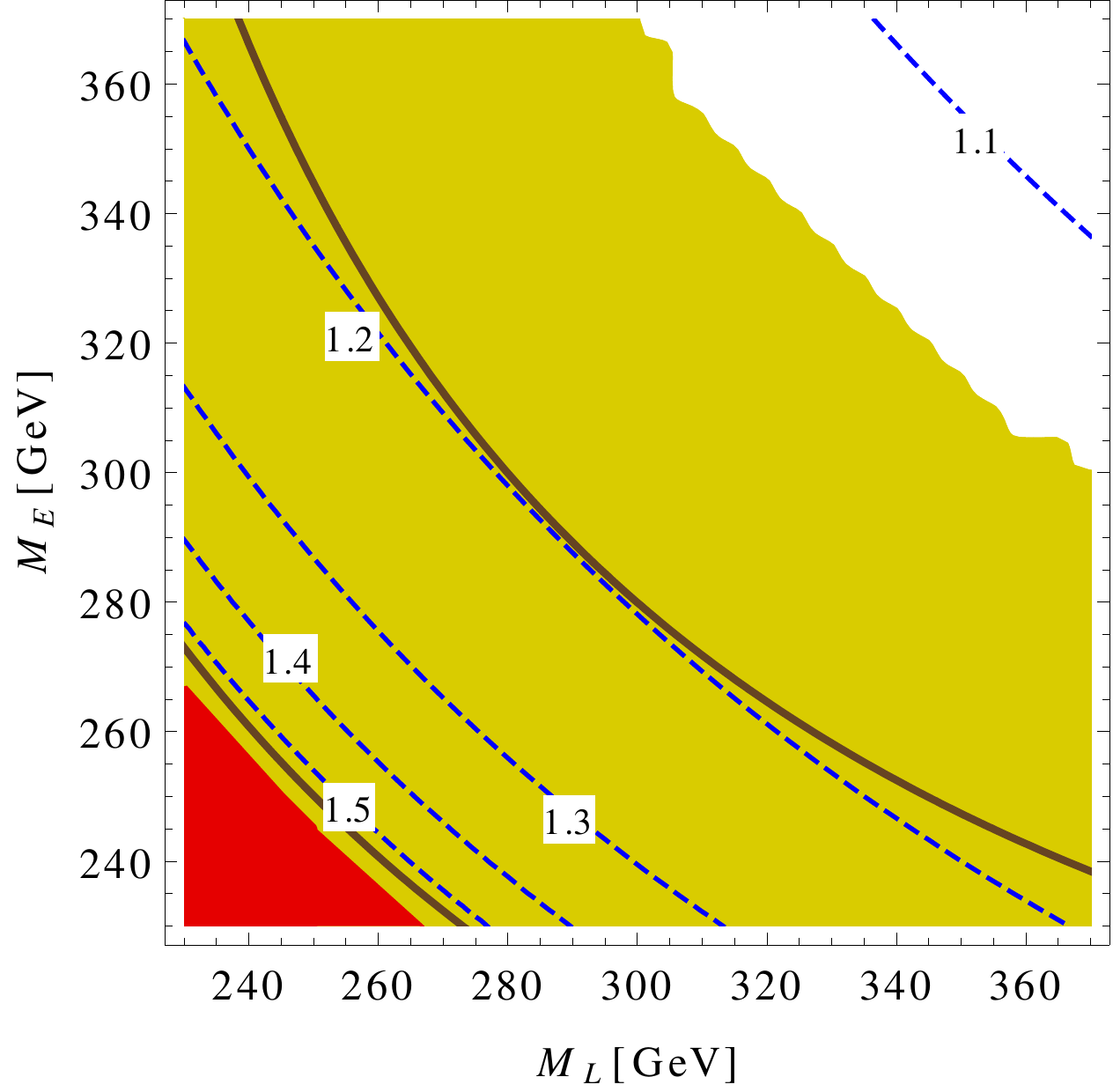}
\includegraphics[width=.47\textwidth]{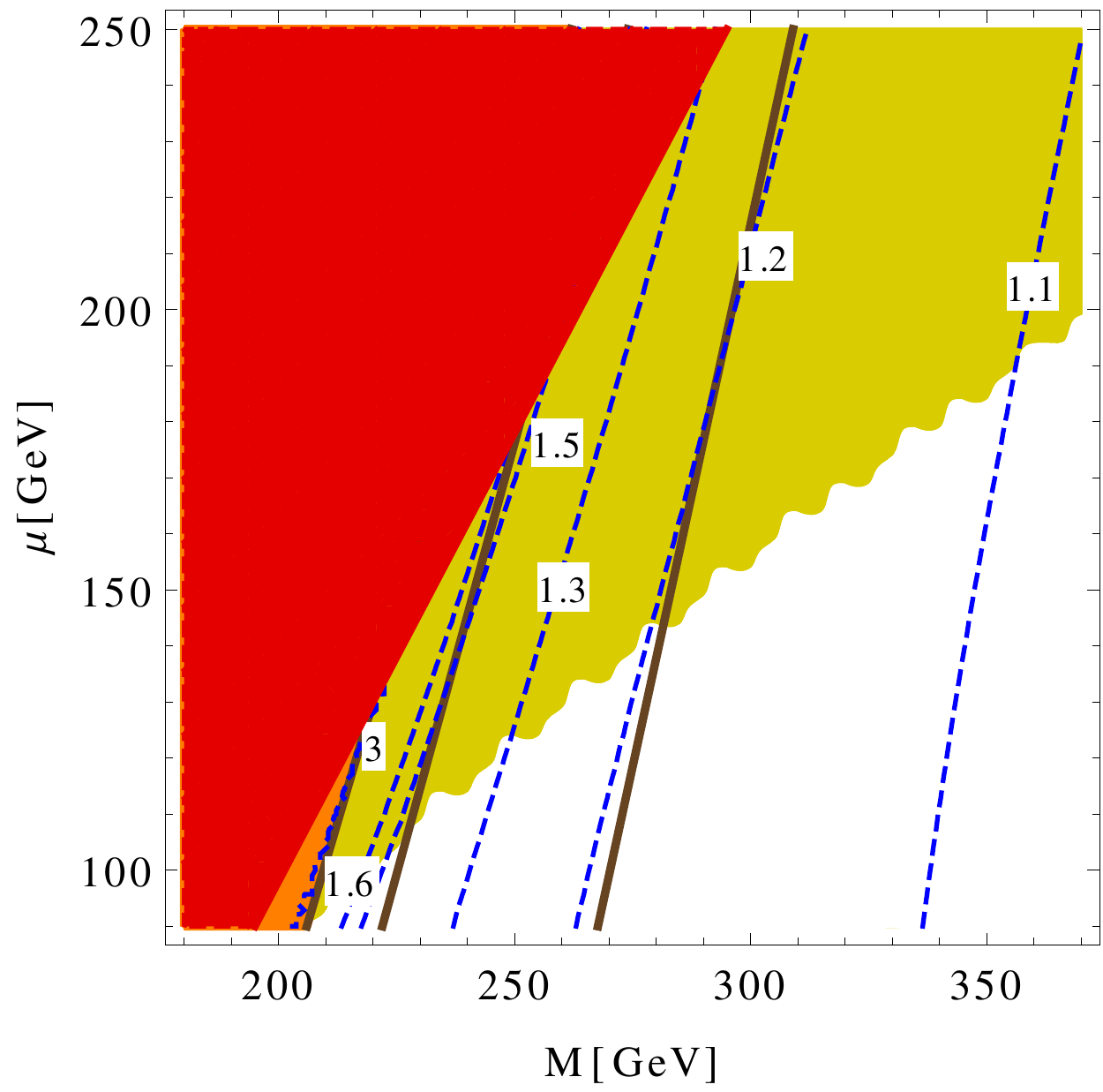}\hspace*{.5cm}
\caption{Same as Fig.~\ref{fig:hightan} for $\tan\beta=2$. In the left panel $\mu=175$~GeV.  
}
\label{fig:lowtan}
\end{figure}

In Fig.~\ref{fig:lowtan} we show the diphoton rates for $\tan\beta = 2$, while all other parameters are the same as before. Here we expect that the fermions contribute up to 30\% enhancement to the diphoton rate, such that higher total rates should be possible. However it turns out that the vacuum stability constraint forces us to consider lower values of $\mu$, such that the overall enhancement is not significantly higher than in the large $\tan\beta$ case. 

More precisely, it can be seen in the left panel that an enhancement of over 50\% is possible with $\mu$ as low as 175~GeV and $m_{\tilde{E}_1}> 90$~GeV, which is not possible in the large $\tan\beta$ case. However the right plot clearly shows that larger values of $\mu$ do not further increase $R_{\gamma\gamma}$, since the stability constraint requires larger values of $M$ at the same time. Moving along the $m_{\tilde{E}_1} = 90$~GeV contour, the ratio of fermionic to scalar contributions, $\Delta_{E}/\Delta_{\tilde{E}}$, decreases from about 0.5 at $\mu = 120$~GeV to 0.3 at $\mu=175$~GeV. 
Enhancements of order 100\% are again only possible if we lower the slepton masses below the conservative LEP bound. 

\

Since the soft mass terms are set to zero, the vector masses $M_L$ and $M_E$ are needed to lift the sleptons above the LEP bound. We have seen in Sec.~\ref{sec:veclep} that the leptonic contribution is maximized around $M_L = M_E = 200$~GeV. However such low values are not allowed since the mass of the lightest slepton would drop below the LEP limit. It is therefore interesting to explore what happens when we add small soft terms to stabilize the slepton masses, such that both the leptonic and the slepton contributions to $R_{\gamma\gamma}$ can be maximized. 

It is instructive to qualitatively discuss the impact of the various elements of the slepton mass matrix before proceeding to finding the maximum possible $R_{\gamma\gamma}$ in the presence of soft terms. The main role of the off-diagonal terms is to increase the difference between the eigenvalues, while the sum of the eigenvalues is given by the trace of the mass matrix. Therefore, the parameters that only appear in off-diagonal terms increase the difference between eigenvalues keeping the sum constant, effectively resulting in lowering the lowest eigenvalue.
\begin{figure}
\center
\includegraphics[width=.47\textwidth]{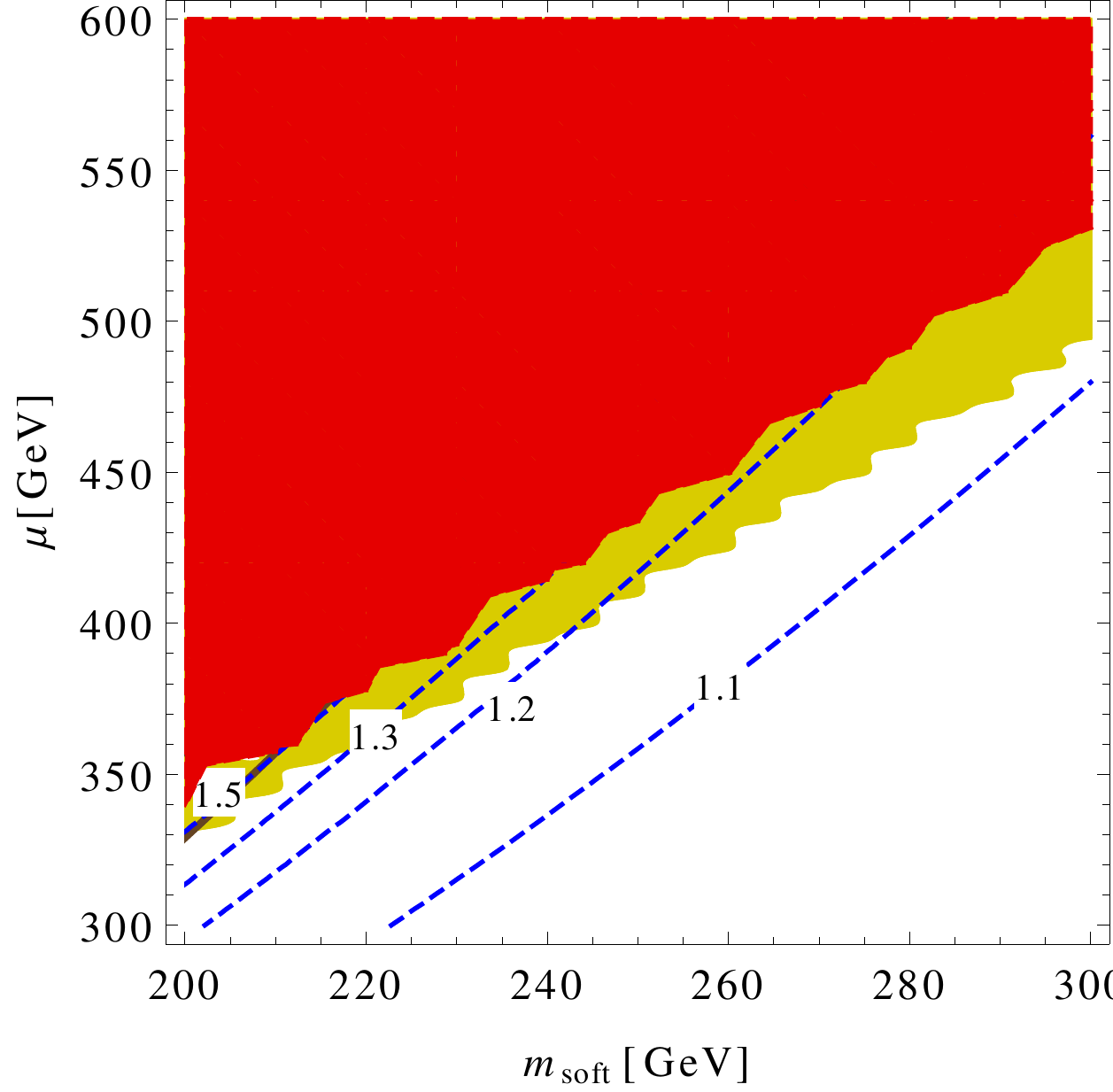}
\includegraphics[width=.47\textwidth]{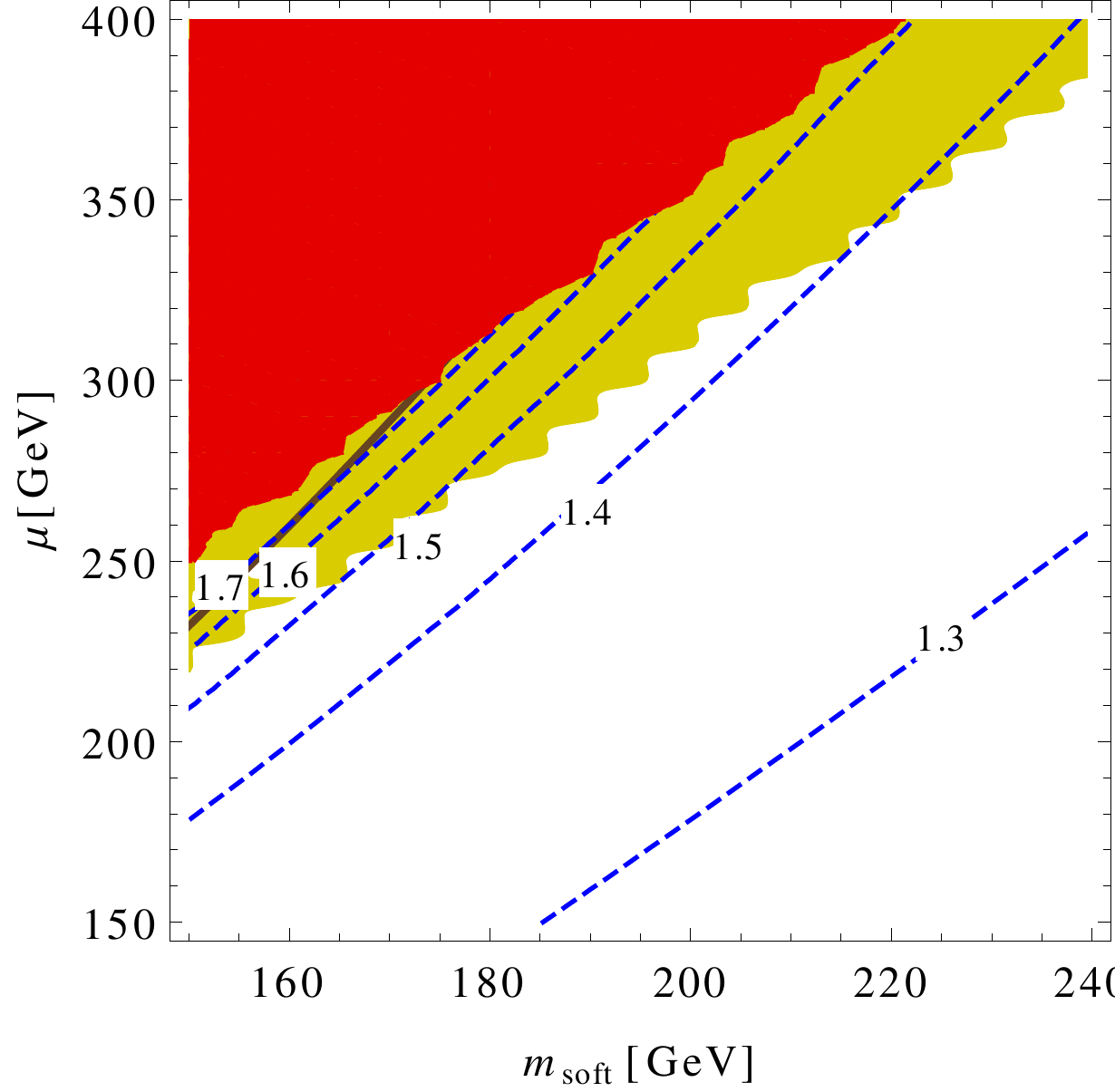}\hspace*{.5cm}
\caption{Diphoton rates with nonzero soft mass terms for the sleptons, $m_i^2 =m_{\rm soft}^2$. The vector mass terms $M_L=M_E=M$ are chosen such that the lightest lepton has a mass around 100~GeV. The left panel shows the result for $\tan\beta=60$ and $M=150$~GeV, while the right panel is for $\tan\beta=2$ and $M=190$~GeV. 
Colors same as in Fig.~\ref{fig:hightan}.
}
\label{fig:crazy}
\end{figure}

It is clear from the structure of the slepton mass matrix given by Eq.~(\ref{eqn:slepmass}) that the $\mu$ parameter and the bilinear holomorphic soft terms are only present in the off-diagonal terms. Thus both of these can drive the lightest slepton mass below the LEP bound as well as destabilize the vacuum. On the other hand, as illustrated in Eq.~(\ref{eqn:prefactor1}), these off-diagonal terms are essential to produce $R_{\gamma\gamma}$ significantly higher than 1. These two parameters have similar impact on $R_{\gamma\gamma}$, lightest slepton mass and stability. Therefore, in order to find the maximum enhancement, we set the holomorphic soft terms to zero and achieve the $h\rightarrow\gamma\gamma$ enhancement only through $\mu$ as before. 

The non-holomorphic soft terms $m_i^2$ that appear on the diagonal of~(\ref{eqn:slepmass}) can then be used to lift the lightest slepton mass above the LEP limit. For simplicity we will take them to be equal, $m_i^2 = m_{\rm soft}^2$. Fig.~\ref{fig:crazy} shows the enhancement rates that can be obtained for $\tan\beta=60$ (left) and $\tan\beta=2$ (right). 
The vector mass scale $M$ is chosen such that the lightest lepton mass is about $100$~GeV, corresponding to $M=150$~GeV ($\tan\beta=2$) and $M=190$~GeV ($\tan\beta=60$), respectively. 
In the case of large $\tan\beta$, we note that an enhancement of 50\% is now possible without going below the conservative LEP limit on the slepton masses, an improvement of about 10\% compared to the case without soft terms. Similarly, in the low $\tan\beta$ case, we can now get up to 75\% enhancement of the diphoton rate without making the sleptons dangerously light. Furthermore one should note that the absolute stability limit is less constraining now, and values of $R_{\gamma\gamma}$ of 1.5 and 1.6 are compatible with absolute stability for high and low $\tan\beta$, respectively.

\subsection{Split sleptons}\label{sec:ssl}
In Sec.~\ref{sec:vacuum} we have seen that imposing absolute vacuum stability puts modest constraints on the parameter space when only two charged sleptons are allowed to get a VeV, but that the constraints become  strong when all four sleptons are taken into consideration. 

Here we will consider a scenario where the sleptons are split by TeV scale soft masses $m_{\tilde{E}_i}$ for the double primed fields $\tilde{e}_L''$ and $\tilde{e}_R''$. This will improve the vacuum stability, since non-zero expectation values for these fields are unlikely to give vacua deeper than the EWSB vacuum. 
 The remaining light charged degrees of freedom are the two sleptons $\tilde{e}_L'$ and $\tilde{e}_L'$ as well as both the charged leptons. Phenomenologically this is a combination of the vector-like lepton model~\cite{ASW} with the light stau scenario~\cite{Carena:2011aa}. 

\begin{figure}
\center
\includegraphics[width=7cm]{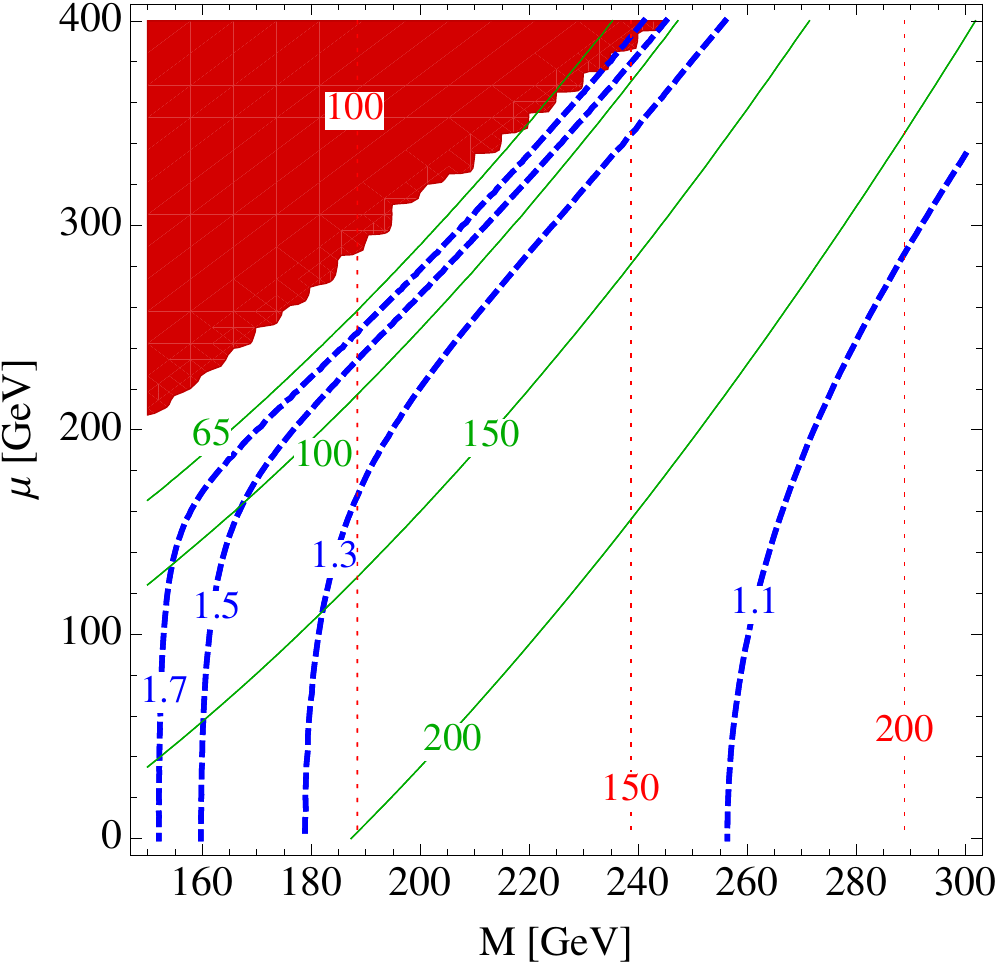}\hspace*{.4cm}
\includegraphics[width=7cm]{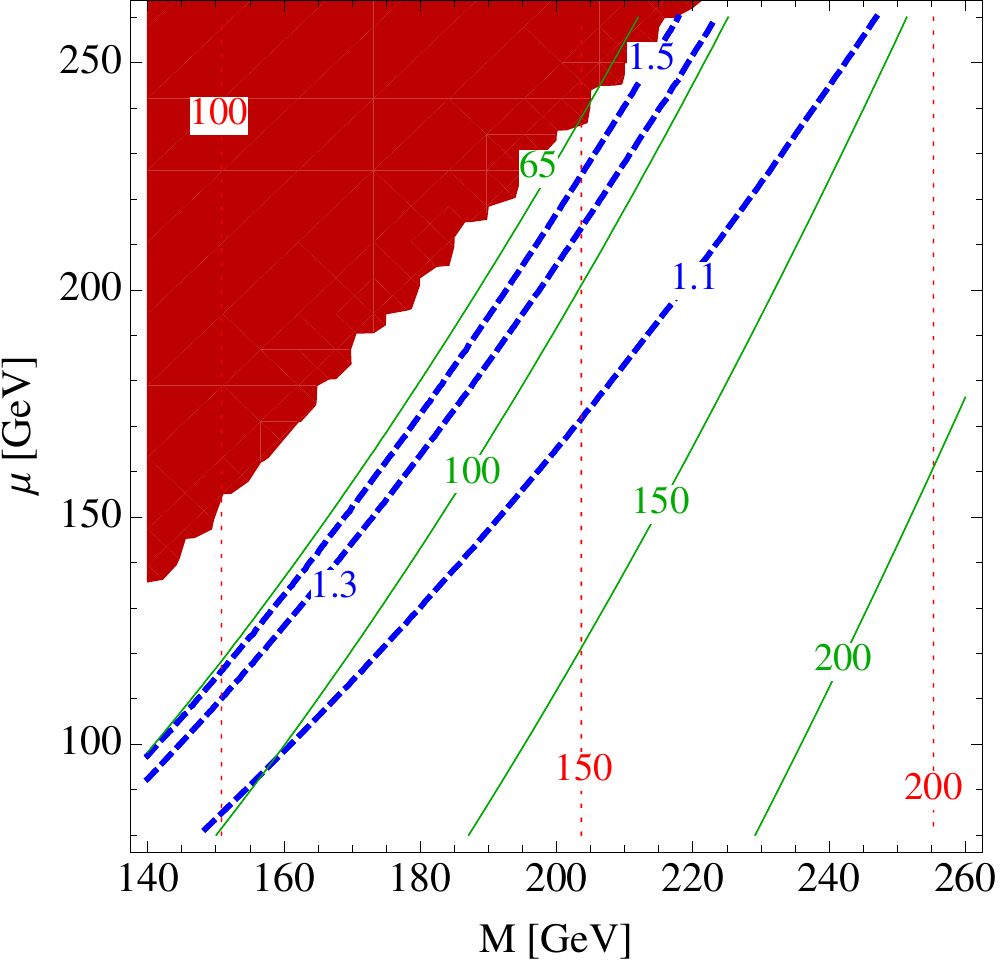}
\caption{Enhancement of the $h\to\gamma\gamma$ rate in the split slepton scenario for $\tan\beta = 2$ (left) and $\tan\beta=30$ (right), with charged Yukawa couplings $y'_c = 0.9$ and $y''_c=0.7$. Shown are iso-contours of $R_{\gamma\gamma}$ (blue, dashed), of the lightest slepton mass (green, solid) and of the lightest lepton mass (red, dotted). Note that values of $\mu$ below $100$~GeV are in conflict with limits on chargino masses. The red shaded area indicates either a meta-stable or unstable vacuum. We do not differentiate between the two here since both regions lie outside of the LEP allowed region. 
}
\label{fig:sl1}
\end{figure}

Absolute vacuum stability is guaranteed provided that the inequality~(\ref{eqn:mubound}) is satisfied. As can be seen from Fig.~\ref{fig:analyticResult} the conservative LEP limit on the mass of the lightest charged particle, $m_{\tilde{E}_1} \gtrsim 90$~GeV, is in general more constraining, so that the vacuum stability constraint is automatically satisfied for most phenomenologically viable parameter points. Regions of parameter space that are not absolutely stable will be indicated in the plots, but one should keep in mind that they can still be phenomenologically viable if the much weaker meta-stability bound is satisfied. 

We can again distinguish two scenarios, low $\tan\beta$, where leptons and the light sleptons contribute to $R_{\gamma\gamma}$, and the large $\tan\beta$ regime, where only the two light sleptons can make a notable contribution.

Let us first consider the small $\tan\beta$ case. For definiteness we take $\tan\beta=2$ and, as in the previous sections, $y_c'=0.9$ and $y_c''=0.7$. In Fig.~\ref{fig:sl1} (left) we show $R_{\gamma\gamma}$ in the $\mu$--$M$ plane, where $M=M_L = M_E$ is the vector mass scale. As can be seen, the enhancement can be increased both by lowering the lightest slepton mass (i.e. increasing $\mu$) or by lowering the lightest lepton mass (decreasing $M$). When both masses are close to 100~GeV an enhancement of about 40\% can be obtained. Values up to 70\% can be reached by lowering the lightest slepton mass further, while still being consistent with limits from the LEP experiments and with absolute vacuum stability. 

When $\tan\beta$ is increased, the leptonic contributions are suppressed. The case of $\tan\beta=30$ is shown in Fig.~\ref{fig:sl1} (right). $R_{\gamma\gamma}$ is now independent of the lightest lepton mass, and an enhancement of 30\% or more can only be obtained when the lightest slepton mass is below 100~GeV, and only for sufficiently large values of $\mu$. Imposing absolute vacuum stability is more constraining here, which disfavors the large $\mu$ region where $R_{\gamma\gamma}$ can be of order 1.5 or higher. 

\

In Secs.~\ref{sec:veclep}-\ref{sec:ssl}, we have focussed on the maximal possible values of $R_{\gamma\gamma}$ that can be obtained in each scenario. In contrast to many existing models, we find that at least in some scenarios it is possible to obtain enhancements of more than 50\% while at the same time the new particle masses can be kept above the conservative LEP bound, vacuum stability is maintained and all couplings remain perturbative up to high scales. Nevertheless it is evident from our figures that some amount of tuning of the parameters is necessary to obtain such large values for $R_{\gamma\gamma}$, while more modest enhancements of $20\%-40\%$ are possible in larger regions of parameter space and thus appear more natural. Such an enhancement is well in agreement with the signal strength indicated by the combination of the updated ATLAS and CMS results	~\cite{Chatrchyan:moriond,ATLASmoriond}. 

%%%%%%%%%%%%%%%%%%
\section{Conclusions}\label{sec:conclusions}
The recent discovery of a Higgs-like particle at the LHC opens a new era in particle physics. A very important task is to study
the properties of this particle in detail, and to analyze any possible deviation from the SM predictions that might signal the presence of new physics. 
Currently, the measured Higgs-induced diphoton production rate is 2.3~$\sigma$ above the SM 
prediction at the ATLAS experiment~\cite{ATLASmoriond}.  This provides a motivation for the study of new physics scenarios that can lead to an enhancement of the 
diphoton decay width.   Vector-like leptons provide an extension 
of the SM that leads to such an enhancement.

Quite generally, the presence of new weakly interacting particles with strong couplings to the Higgs boson can provide an enhancement of the loop-induced diphoton rate, but also lead to the presence of new vacua deeper than the physical one. In the case of vector-like leptons such vacua arise at large values of the Higgs fields due to the evolution of the quartic coupling of the Higgs to negative values.  Enhancements of the diphoton rate to values larger than 1.5 times the SM one can only be obtained if new physics stabilizes the vacuum at scales smaller than a few TeV. Supersymmetry provides a natural extension of this model in which the vacuum of the Higgs sector is stabilized by the contributions of sleptons.

In this article we study the supersymmetric extension of the vector-like lepton theory introduced in Ref.~\cite{ASW}.  We showed the inclusion of a whole vector-like family can lead to a unified theory with values of the gauge couplings that are close to the non-perturbative bound at scales close to the GUT scale.  In order to enhance the effects on the Higgs diphoton decay rate, we chose values of the Yukawa couplings leading to large values at the GUT scale, but still consistent with a perturbative treatment of the theory.  Vector-like squarks are considered to be heavy and therefore have an impact only in the determination of the Higgs mass. 

The phenomenological properties of this supersymmetric extension depends strongly on the values of the soft breaking parameters. For large values of the scalar soft supersymmetry breaking parameters, the theory at low energies reduced to the one studied before. However, for the same values of the Yukawa couplings, the lepton contributions are suppressed by a $\sin 2 \beta/2$ factor and, together with modified values of the perturbativity bounds on these couplings with respect to the SM ones, the enhancements of the diphoton rate remain smaller than about 30 percent for lepton masses above 100~GeV.

For small values of the scalar soft supersymmetry-breaking parameters, the main source of supersymmetry breaking in the Higgs-slepton potential is provided by the Higgsino mass parameter $\mu$.  For light sleptons, large values of $\mu$ tend to induce new charge breaking minima in the spectrum, and therefore in the region consistent with vacuum stability light sleptons are associated with relatively light leptons. It follows that both the fermion and the scalar lepton contributions to the Higgs-induced diphoton production rate tend to be important, except for the large $\tan\beta$ regime where the lepton contributions decouple.
We find that  for lepton and slepton masses larger than 100~GeV  enhancements of order 50 and 40 percent may be obtained for small and larger values of $\tan\beta$ ($\tan\beta = 2$ and 60), respectively.  More generally, since the LEP bound depends on the value of the neutrino mass parameters, one can consider leptons and sleptons as light as 62.5~GeV, for which much larger enhancements of the Higgs diphoton rate may be obtained. 

Finally, we considered a split slepton scenario, in which the soft supersymmetry-breaking parameters of the new right-handed leptons are considered to be large, while the left-handed ones are kept small.  In such a case, the theory is similar to the light-stau scenario, but the lepton contributions remain relevant. We showed that in such a case enhancements of the Higgs-induced diphoton rate of the order of 50 percent and 30 percent can be obtained for small and large values of $\tan\beta$ for lepton and slepton masses above 100~GeV, while as before larger values may be obtained if these bounds were relaxed.  

In order to avoid flavor problems, we have introduced a parity symmetry under which the new states are odd. If in addition R-parity is conserved this guarantees three stable particles: the ordinary MSSM LSP as well as the lightest parity odd leptons and sleptons. Assuming that in each sector a neutral particle is the lightest state, this leads to a scenario with multicomponent dark matter with possibly interesting phenomenological consequences. 

Even in the light of the recently presented CMS results~\cite{Chatrchyan:moriond}, there are large regions of parameter space in which the vector like leptons and sleptons are present at the weak scale and remain compatible with data. The masses of the new particles in such region can be small enough to make them interesting to study from the perspective of dark matter and low-scale leptogenesis. They can also be probed at the LHC, what  serves as another motivation to study supersymmetric vector-like leptons in detail.

\acknowledgments{We would like to thank C.~Wainwright for discussions. Work at ANL is supported in part by the U.S. Department of Energy, Division of High Energy Physics, under grant number DE-AC02-06CH11357, at EFI under grant number DE-FG02-90ER-40560, and at UIC under grant number DE-FG02-12ER41811.}

\paragraph*{Note added:} While finalizing the manuscript for submission, Ref.~\cite{Feng:2013mea} appeared, which also considers a supersymmetric extension for the vector-like lepton scenario. Different from the present work, the authors do not impose perturbativity of the Yukawa couplings at the GUT scale, and therefore allow larger Yukawa couplings at the electroweak scale. We also have shown that the relevance of vacuum stability constraints depends on the choice of soft breaking parameters.

%%%%%%%%%%%%%%%%
%%%%%%%%%%%%%%%%

\end{document}